\documentclass[3p]{elsarticle}
\pdfoutput=1
\usepackage[utf8]{inputenc}
\usepackage{hyperref,amsmath,amsfonts,natbib,graphicx,bm,subfig,floatrow,multirow,wrapfig}
\usepackage[ruled,vlined]{algorithm2e}
\usepackage[dvipsnames]{xcolor}

\newcommand{\bu}{\ensuremath{\mathbf{u}}}
\newcommand{\bx}{\ensuremath{\mathbf{x}}}
\newcommand{\by}{\ensuremath{\mathbf{y}}}

\newcommand{\pder}[2]{\ensuremath{\frac{\partial #1}{\partial #2}}}
\newcommand{\spder}[2]{\ensuremath{\frac{\partial^2 #1}{\partial #2^2}}}

\graphicspath{{figures/}}

\journal{Computers and Fluids}

\SetKwBlock{DummyBlock}{}{}

\begin{document}
\begin{frontmatter}

\title{Gradient-enhanced stochastic optimization of high-fidelity simulations}

\author[sorbonne]{Alejandro Quirós Rodríguez\corref{corresponding}}
\cortext[corresponding]{Corresponding author}
\ead{alejandro.quiros_rodriguez@etu.sorbonne-universite.fr}

\author[esi]{Miguel Fosas de Pando}
\ead{miguel.fosas@uca.es}

\author[sorbonne]{Taraneh Sayadi}
\ead{taraneh.sayadi@sorbonne-universite.fr}

\address[sorbonne]{Sorbonne Université, Institut Jean Le Rond $\partial$'Alembert, UMR 7190, 4 Place Jussieu, 75252 Paris Cedex 05, France}
\address[esi]{Dpto.\ Ing.\ Mecánica y Diseño Industrial, Escuela Superior de Ingeniería, Universidad de Cádiz, Av.\ de la Universidad de Cádiz 10, 11519 Puerto Real, España}

\begin{abstract}
Optimization and control of complex unsteady flows remains an important challenge due to the large cost of performing a function evaluation, i.e. a full computational fluid dynamics (CFD) simulation. Reducing the number of required function evaluations would help to decrease the computational cost of the overall optimization procedure. In this article, we consider the stochastic derivative-free surrogate-model based Dynamic COordinate search using Response Surfaces (DYCORS) algorithm \cite{regis_combining_2013} and propose several enhancements: First, the gradient information is added to the surrogate model to improve its accuracy and enhance the convergence rate of the algorithm. Second, the internal parameters of the radial basis function employed to generate the surrogate model are optimized by minimizing the leave-one-out error in the case of the original algorithm and by using the gradient information in the case of the gradient-enhanced version. We apply the resulting optimization algorithm to the minimization of the total pressure loss through a linear cascade of blades, and we compare the results obtained with the stochastic algorithms at different Reynolds numbers with a gradient-based optimization algorithm. The results show that stochastic optimization outperforms gradient-based optimization even at very low $Re$ numbers, and that the proposed gradient-enhanced version improves the convergence rate of the original algorithm. An open-source implementation of the gradient-enhanced version of the algorithm is available in  \cite{quiros_rodriguez_dycors_2022}.
\end{abstract}

\begin{keyword}
Stochastic optimization, Surrogate models, Radial Basis Functions, Gradient-enhanced Radial Basis Functions, High-fidelity simulations
\end{keyword}
\end{frontmatter}

\section{Introduction}

Progress in computational capabilities during the past decades have allowed computational fluid dynamics (CFD) to become an ever more present tool in the description and the prediction of complex unsteady flows. However, the computational cost associated with such high-fidelity simulations precludes them from being routinely used in state-of-the-art optimization algorithms, without resorting to reduced order models. Thus, the development of strategies that reduce the number of function evaluations, i.e. CFD simulations, in such optimization algorithms is crucial to achieve an acceptable computational cost.

Optimization algorithms generally fall under two main categories: (i) gradient-based, or (ii) derivative-free methods. Gradient methods rely on the value of local derivatives to identify a descent direction. This derivative is most commonly calculated using analytical expressions or finite differences. Both strategies are inapplicable to high-fidelity simulations: analytical expressions are usually not available and finite difference becomes very expensive in the case of unsteady high-fidelity simulations, and is susceptible to noise. Alternatively, gradient information can be extracted using adjoint-based algorithms \cite{pironneau_optimum_1974}. Adjoint-based optimization has been widely used in fluid mechanics, from areas dominated by linear dynamics (e.g. acoustics and thermo-acoustics \cite{jameson_optimum_1998,juniper_triggering_2011}), to nonlinear systems (e.g. analysis of high-lift airfoils, mixing  enhancement and minimal seeds for transition to turbulence \cite{foures_optimal_2014,schmidt_three-dimensional_2013,rabin_designing_2014}). Recently, their application to more complex flow regimes, such as reactive and interfacial flows have also been investigated \cite{fikl_control_2020,hassan_sensitivity_2021,hassan_adjoint-based_2021,hassan_adjoint-based_2021-1}. However, as demonstrated by \cite{fikl_control_2020}, the objective function encountered in such flows can have multiple minima, rendering the application of gradient methods difficult. In addition, the presence of turbulence makes the gradient-based approach inadmissible in many complex flow scenarios. Derivative-free methods elevate these challenges and have been applied successfully to optimization in fluid mechanics~\cite{marsden_computational_2008,pierret_multidisciplinary_2006}. Their main drawback, however, is the requirement for many function evaluations, which proves to be too costly when dealing with cases of practical interest.

Due to these disadvantages, the application of these methods to optimization problems involving high-fidelity unsteady simulations is not straightforward. A suitable alternative for cases with expensive function evaluations is one that is based on a response surface model (also known as a surrogate model or a meta-model), which is, in essence, an inexpensive approximate model of the underlying expensive function. Performing the optimization procedure on a surrogate surface greatly reduces the number of calls to the expensive high-fidelity model. Surrogate model optimization has been used extensively to identify promising points for function evaluations \cite{jones_taxonomy_2001,gutmann_radial_2001,moore_memory-based_1995,powell_trust_2003} using different interpolation techniques that have been proposed, e.g. Least Squares (LS) \cite{myers_response_2009}, Kriging \cite{daya_sagar_fifty_2018}, Radial Basis Functions (RBF) \cite{powell_theory_1992} and Support Vector Regression (SVR) \cite{smola_tutorial_2004}. The most promising point on the surrogate model can be determined by several techniques, such as the Adaptive Response Surface Method (ARSM) \cite{gary_wang_adaptive_2001}, Efficient Global Optimization (EGO) \cite{jones_efficient_1998} and DYCORS \cite{regis_combining_2013}, to name a few.

Although surrogate model optimization reduces the number of expensive function calls dramatically, it still suffers from the curse of dimensionality, especially when the number of design variables increases \cite{han_weighted_2017}. In addition, typical algorithms still require a large number of function evaluations to be applicable to the problems of practical interest. In order to ameliorate these restrictions, gradient information can be incorporated into the surrogate model. Two main approaches are (i) constructing the surrogate surface using the gradient as well as the local function value \cite{leary_derivative_2004,march_gradient-based_2011,noel_new_2012} or (ii) using multiple start algorithms \cite{peri_multistart_2012,ugray_scatter_2007}. Both approaches show promising results, suggesting that a judicious combination of derivative-free and gradient-based methods can lead to an efficient procedure that converges to the global minimum with a limited number of expensive function evaluations. 

In this study, the DYCORS algorithm~\cite{regis_stochastic_2007} is adopted as the basis of the surrogate model optimization procedure. This algorithm is particularly attractive due to its fast convergence to the global minimum in a high-dimensional parameter space. This characteristic is necessary in applications of interest to CFD, since the control function is most commonly a parametrized/discretized function, distributed in space. To our best knowledge, this work presents one of the first applications of DYCORS to unsteady flows. We aim to provide a measure of its performance at different regimes such as steady, unsteady and non-deterministic flow. In addition, the use of local gradient information is proposed to improve the accuracy of the surrogate model, resulting in a gradient-assisted surrogate model optimization that aims at reducing the number of required function evaluations to reach the global optimum. Moreover, the optimization of the internal parameters of the surrogate model has been included in the optimization procedure to further enhance its accuracy. The resulting optimization algorithm is applied to control the unsteady flow around a linear cascade of compressor rotor blades.

The paper is organized as follows. First, in Section \ref{sec:optimization_framework} a detailed description of the stochastic optimization algorithm is provided and the enhancements to the original algorithm are highlighted. Then, its performance is assessed in the context of numerical flow simulations. The governing equations and the numerical schemes of the underlying flow solver are briefly presented in Section \ref{sec:equations}. In Section \ref{sec:results}, an application of this algorithm to the reduction of total pressure loss through a linear cascade of blades is presented and the results are discussed. Finally, we provide in Section \ref{sec:conclusions} concluding remarks and suggestions for future work.

\section{Optimization framework}
\label{sec:optimization_framework}
\begin{figure}[hp]
    \centering
    \subfloat[][The algorithm is initialized with $m=6$ initial sampling points distributed through the domain by means of a Latin Hypercube Sampling technique. The objective function is evaluated at these points.]{\includegraphics[width=0.45\textwidth]{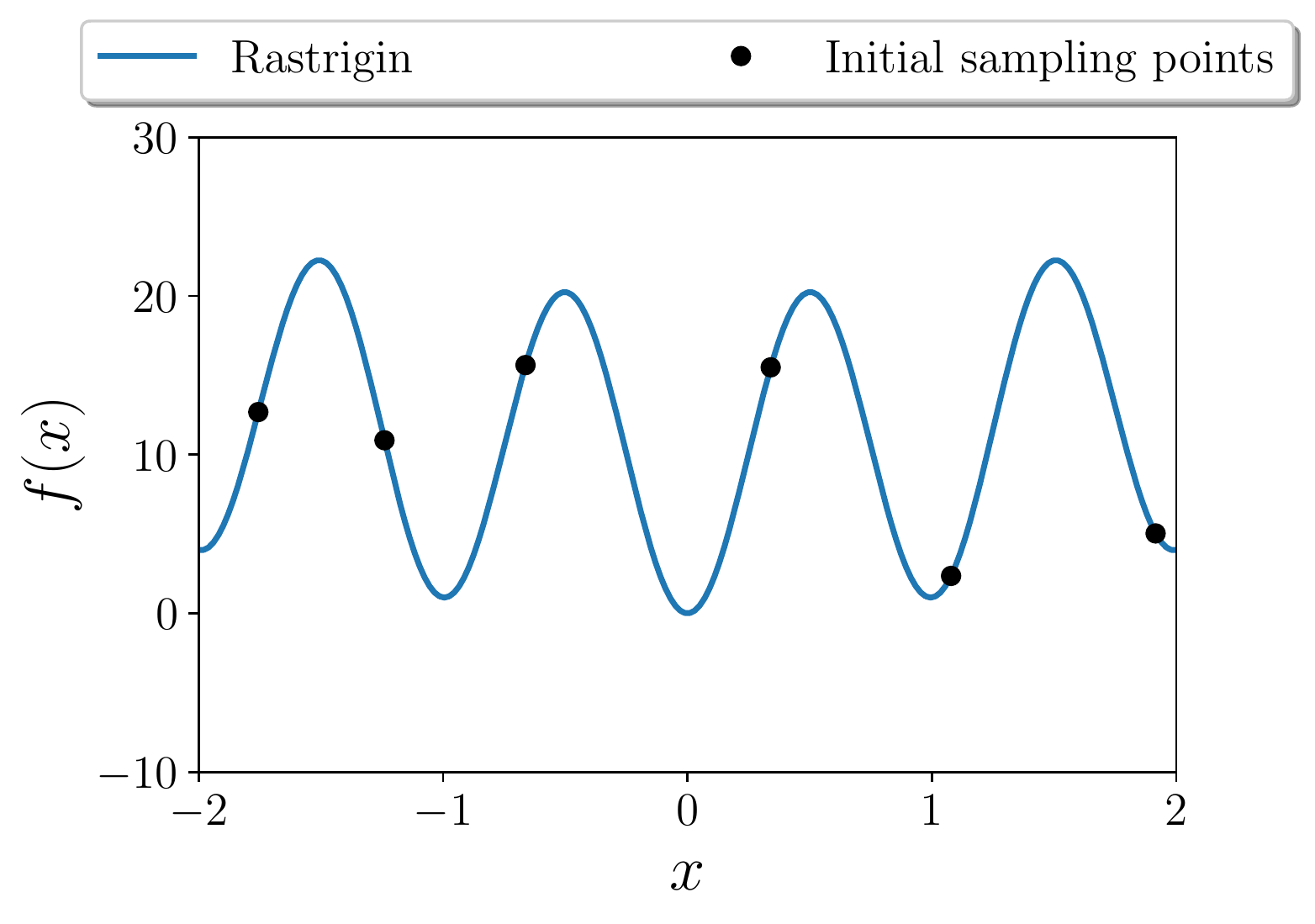} }\hfil
    \subfloat[][The surrogate model is built using previously evaluated points. This surrogate model is then evaluated at $k=200$ trial points (shown in orange) drawn from a normal distribution centered at current best point (PDF shown in gray).]{\includegraphics[width=0.45\textwidth]{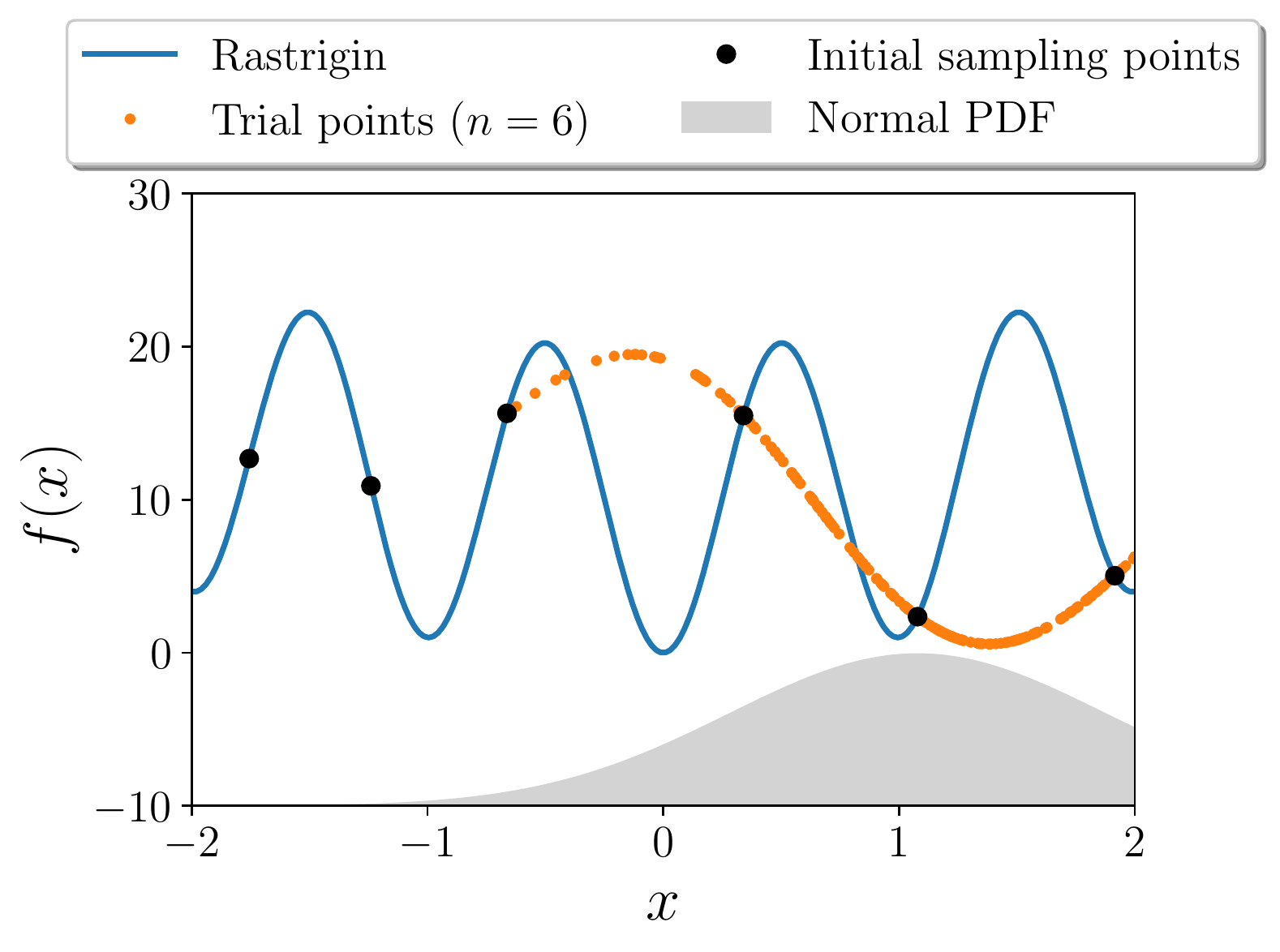} } \\
    \subfloat[][The most promising point (shown in red) is selected based on a given criteria.]{\includegraphics[width=0.45\textwidth]{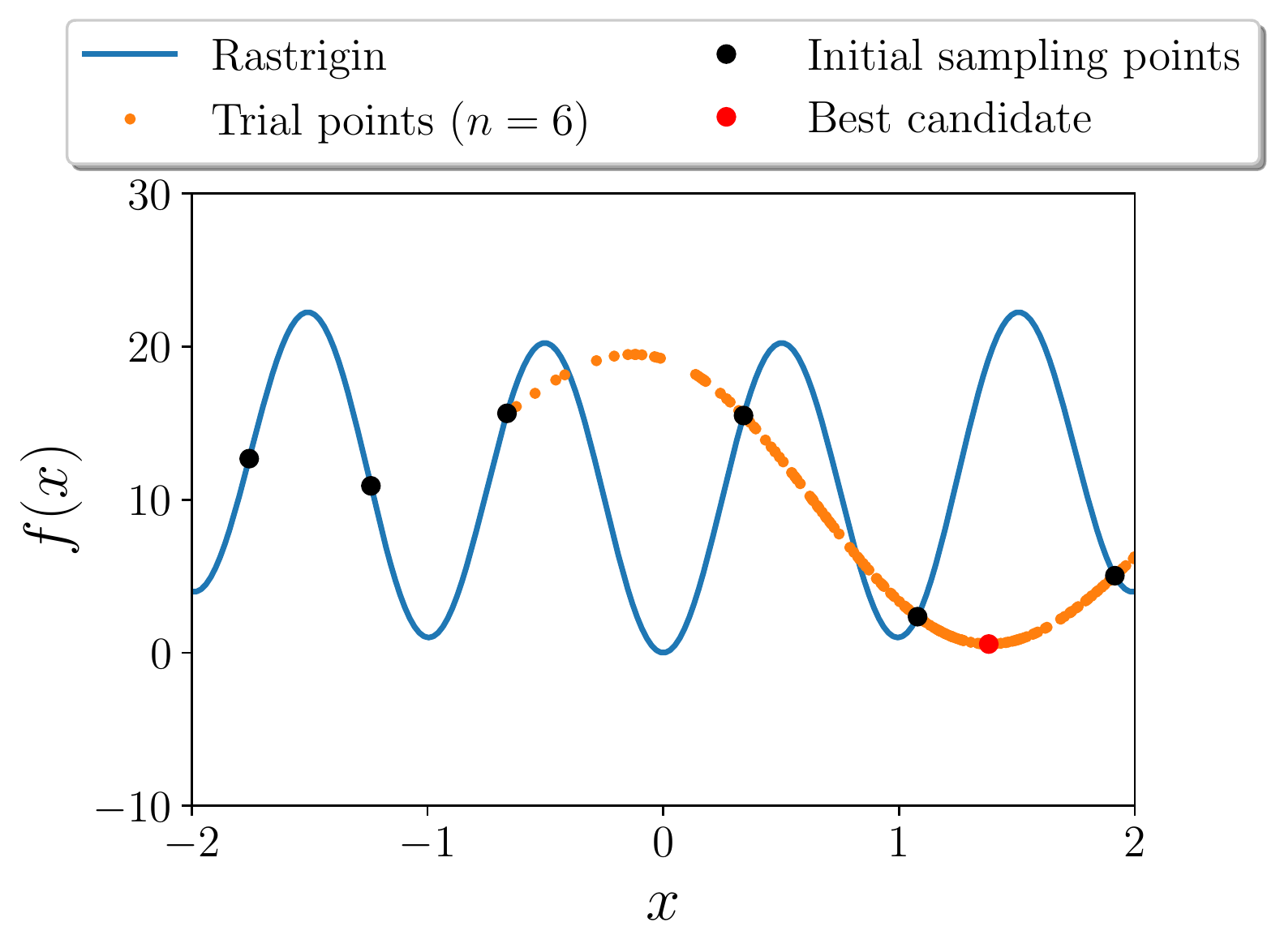} }\hfil
    \subfloat[][The objective function is evaluated at the best candidate point and the surrogate model is updated. In this case, the value of the objective function at the best candidate point does not improve the previous best, and the normal distribution used to generate the new trial points is centered at the same location as the previous iteration.]{\includegraphics[width=0.45\textwidth]{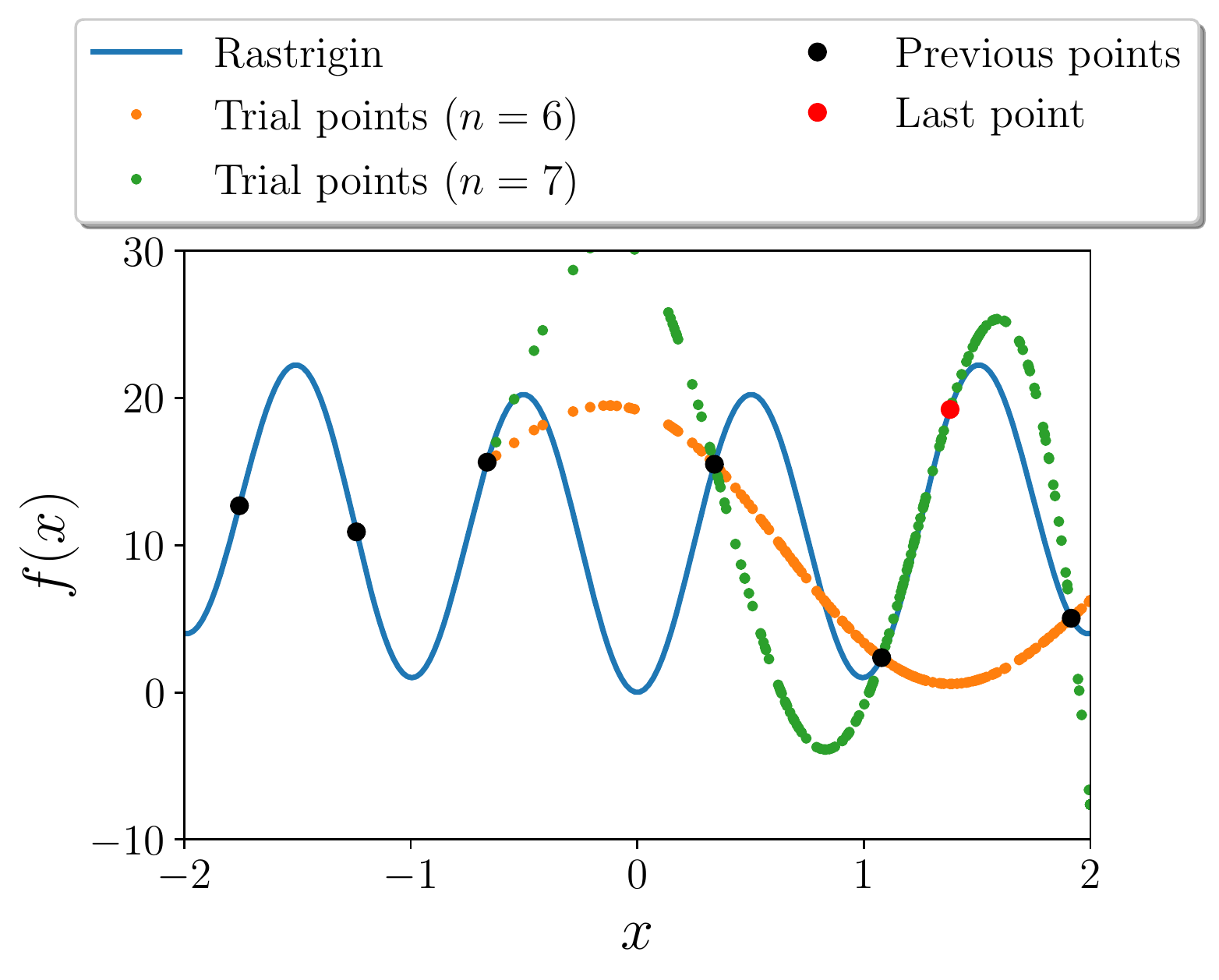} } 
    \caption{Illustration of the main steps performed during one iteration of DYCORS algorithm. The one-dimensional Rastrigin function is used as the objective function defined on the domain $x \in [-2,2]$.}
    \label{fig:example_optim}
\end{figure}

The Dynamic Coordinate Search using Response Surfaces (DYCORS) algorithm developed in \cite{regis_stochastic_2007} is first described in this section, and then extended to include derivative information. This algorithm is chosen owing to its performance in a high-dimensional parameter space. Once the algorithm is initialized by evaluating the objective function at selected initial sampling points, it produces a sequence of candidate solutions until a stop criterion is met. At each iteration, the following operations are performed: 
\begin{itemize}
    \item Construction of the surrogate model using information from previously-evaluated points, Fig.~\ref{fig:example_optim}(a). 
    \item Generation of trial points and evaluation using the surrogate model, Fig.~\ref{fig:example_optim}(b)
    \item Selection of best candidate point among the trial points, Fig.~\ref{fig:example_optim}(c). 
    \item Evaluation of the objective function at the best candidate point, Fig.~\ref{fig:example_optim}(d). 
\end{itemize}
This procedure is illustrated in Fig.~\ref{fig:example_optim}, where the one-dimensional Rastrigin function~\cite{rudolph_globale_1990}, a commonly used function to benchmark algorithms in the presence of a large number of local minima, is considered. These steps will be presented below in more detail.

\subsection{Construction of the surrogate model}
\label{sec:surrogate}
In the following, the interpolation technique underlying the surrogate model in the DYCORS algorithm is briefly discussed. This procedure is then modified to include gradient information and a new criterion is introduced to determine the internal parameters of the interpolant.

\subsubsection{Radial Basis Function interpolation}
\label{sec:RBF}
The DYCORS algorithm relies on Radial Basis Functions (RBF) to build the surrogate model. Consider an objective function $f: \mathbb{R}^d \rightarrow \mathbb{R}$, where $d$ is the number of parameters. Taking a set of $n$ points in the parameter space $\bx_1,\dots,\bx_{n} \in \mathbb{R}^d$ and the corresponding values of the objective function  $f(\bx_1),\dots,f(\bx_{n})$, the value of the objective function at a point $\mathbf{y}$ can be approximated by the RBF interpolant~\cite{powell_theory_1992}
\begin{equation} \label{eq:RBF}
    s_n(\by; \bx, \boldsymbol{\lambda}, \mathbf{l}) = \sum_{i=1}^{n} \lambda_i \phi \left ( r(\by, \bx_i, \textbf{l}) \right ),
\end{equation}
where $\phi(\cdot)$ is a kernel function, $\boldsymbol{\lambda}$ is a vector containing the coefficients of the interpolant and $r(\textbf{r}_p, \textbf{r}_c, \textbf{l}) = \left \| \mathrm{diag}(\mathbf{l} ^ {-1}) \left ( \textbf{r}_p-\textbf{r}_c \right ) \right\|$ where $\left \| \cdot \right \|$ is the Euclidean norm, $\textbf{r}_p$ is the point where the radial basis function is going to be evaluated, $\textbf{r}_c$ is the center of the radial function and $\mathbf{l}\in \mathbb{R}^d$ is a vector of internal parameters corresponding to the spatial length-scale of the kernel function in each parameter direction. A wide variety of kernel functions exist, and some of the most popular choices, e.g. the exponential, the Matérn~\cite{matern_spatial_1986} and the cubic kernels, are presented in Table~\ref{tab:kernels}.

\begin{table}
    \centering
    \begin{tabular}{l c}
        Function & Expression \\
        \hline
        Exponential & $\phi(r) = \exp \left( -\dfrac{r^2}{2} \right)$ \\
        Matérn & $\phi(r) = \dfrac{2^{1-\nu}}{\Gamma(\nu)} \left( \sqrt{2\nu} |r|  \right)^\nu K_\nu \left( \sqrt{2\nu}|r| \right)$ \\
        Cubic & $\phi(r) = r^3$
    \end{tabular}
    \caption{Kernel functions, with $r$ being a positive scalar denoting the distance between a point and the center of the radial basis function and $\nu$ an internal parameter of the Matérn kernel referring to the order of the modified Bessel function~$K_\nu$.}
    \label{tab:kernels}
\end{table}

The weights $\boldsymbol{\lambda}$ are determined by setting the value of the interpolant to that of the objective function at every $\mathbf{x}_i$, i.e. $s_n(\mathbf{x}_i; \mathbf{x}, \boldsymbol{\lambda}, \mathbf{l})=f(\mathbf{x}_i)$. However, depending on the kernel choice, the resulting system of equations can be conditionally positive definite~\cite{buhmann_radial_2000}. The interpolant given in Eq.~\ref{eq:RBF} is then modified and polynomials $p$ of degree up to $m$ in $d$ unknowns, i.e. $p \in \Pi^d_m$ are added to the right-hand side; see~\cite{buhmann_radial_2000} for further details. We set $m = 1$ for all the kernels, following \cite{regis_stochastic_2007}. The RBF interpolant then reads
\begin{equation} \label{eq:interp_RBF}
    s_n(\by; \bx, \boldsymbol{\lambda}, \mathbf{l}, \mathbf{c}) = \sum_{i=1}^{n} \lambda_i \phi( r(\by, \bx_i, \textbf{l}) ) + p(\by, \mathbf{c}),
\end{equation}
where $\mathbf{c}= [c^1,\dots,c^{d+1}]^T$ is the vector containing the coefficients of the polynomials. To uniquely determine these coefficients, the above system of equations is augmented by enforcing orthogonality between the coefficients of the kernel functions and the polynomial space $\Pi^d_m$, i.e.
\begin{equation}
    \sum_{i=1}^{n} \lambda_i q_i^j = 0,\quad \mathrm{for} \ j = 1,\dots,d+1,
\end{equation}
where $q_i^1 = 1$ and $q_i^j = x_i^{j-1}$. Finally, the coefficients $\boldsymbol{\lambda}$ and $\mathbf{c}$ are determined by the following linear system
\begin{equation} \label{eq:RBF_system}
    \begin{pmatrix}
        \bm{\mathsf{\Phi}} & \bm{\mathsf{P}} \\ 
        \bm{\mathsf{P}}^T & \bm{\mathsf{0}}
    \end{pmatrix} \begin{pmatrix}
        \boldsymbol{\lambda} \\ 
        \mathbf{c}
    \end{pmatrix} = 
    \begin{pmatrix}
        \mathbf{f} \\ 
        \mathbf{0}
    \end{pmatrix},
\end{equation}
where, $\mathsf{\Phi}_{ij} = \phi(r(\bx_i, \bx_j, \textbf{l}))$ for $i,j=1,\dots,n$ denotes the kernel matrix, $\bm{\mathsf{P}}_i = [1,x_i^1,\dots,x_i^{d}]$ for $i=1,\dots,n$ is the polynomial matrix, and $\mathsf{f}_i = f(\bx_i)$ for $i=1,\dots,n$ is a vector that contains the function values at the evaluated points.

\subsubsection{Gradient-enhanced Radial Basis Function interpolation}
\label{sec:GRBF}
We now turn the attention to Gradient-enhanced Radial Basis Functions (GRBF). The surrogate model can be improved by including local gradient information such that both the function~$f$ and its gradient~$\mathbf{g}$ are matched at the evaluated points. With a more accurate surrogate model, the evaluation of the trial points should provide function values closer to the exact values, thereby, improving the convergence rate of the algorithm. In this case, additional basis functions are introduced to include the local gradient information into the surrogate model. Following \cite{giannakoglou_aerodynamic_2006, laurent_overview_2019, bompard_surrogate_2010}, the interpolation now reads
\begin{equation} \label{eq:interp_GRBF}
    s_n(\by; \bx, \boldsymbol{\lambda}, \textbf{l}, \textbf{c}) = \sum_{i=1}^{n} \lambda_i \phi(r(\by, \bx_i, \textbf{l})) + \sum_{j=1}^{d} \sum_{i=1}^{n} c_i^j \pder{\phi}{r_p^j} \Bigr | _ {r(\by, \bx_i, \textbf{l})},
\end{equation}
where the polynomial term in Eq.\ \eqref{eq:interp_RBF} has been replaced by a term containing the first derivative of the kernel. Note that the size of the vector of coefficients $\mathbf{c}$ is now dependent on the number of evaluated points with a dimension of $nd$, thus, an additional set of equations has to be included to uniquely determine the coefficients. To this end, we differentiate Eq.\ \eqref{eq:interp_GRBF}
\begin{equation} \label{eq:interp_GRBF2}
    \pder{s_n}{r_c^k} \Bigr | _ {(\by; \bx, \boldsymbol{\lambda}, \textbf{l}, \textbf{c})} = \sum_{i=1}^{n} \lambda_i \pder{\phi}{r_c^k} \Bigr | _ {r(\by, \bx_i, \textbf{l})} + \sum_{j=1}^{d} \sum_{i=1}^{n} c_i^j \dfrac{\partial^2 \phi}{\partial r_p^j \partial r_c^k} \Bigr | _ {r(\by, \bx_i, \textbf{l})},
\end{equation}
where both the first and the second derivatives of the kernel function appear. Using Eqs.\ \eqref{eq:interp_GRBF} and \eqref{eq:interp_GRBF2}, the coefficients $\lambda_i$ and $c_i^j$ are determined by the solution of the following linear system
\begin{equation} \label{eq:GRBF}
    \begin{pmatrix}
        \bm{\mathsf{\Phi}} & -\bm{\mathsf{\Phi}}_d \\ 
        \bm{\mathsf{\Phi}}_d^T & \bm{\mathsf{\Phi}}_{dd}
    \end{pmatrix} \begin{pmatrix}
        \boldsymbol{\lambda} \\ 
        \mathbf{c}
    \end{pmatrix} = 
    \begin{pmatrix}
        \mathbf{f} \\ 
        \mathbf{g}
    \end{pmatrix},
\end{equation}
where $\mathbf{g}_i = [ g(\bx_i), \dots, g(\bx_i)]^T$, $i=1,\dots,n$, is the vector with the derivatives of $f$ at the evaluated points. This system is analogous to Eq.\ \eqref{eq:RBF_system} and is guaranteed to be positive definite \cite{laurent_overview_2019}. The matrix with the polynomial terms and the zero matrix of the original RBF formulation have been replaced by the first and the second order derivatives of the kernel matrix $\bm{\mathsf{\Phi}}_d$ and $\bm{\mathsf{\Phi}}_{dd}$, respectively. The derivatives of the kernel matrix can be computed via chain rule,
\begin{align}
    \mathsf{\Phi}_{d_{ij,k}} &= \pder{\phi}{r_c^k} \Bigr | _ {r(\bx_i, \bx_j, \textbf{l})} = \pder{\phi}{r}  \Bigr | _ {r(\bx_i, \bx_j, \textbf{l})} \pder{r}{r_c^k} \Bigr | _ {(\bx_i, \bx_j, \textbf{l})}, \\
    \mathsf{\Phi}_{dd_{ij,kl}} &= \frac{\partial^2 \phi}{\partial r_c^k \partial r_p^l}  \Bigr | _ {r(\bx_i, \bx_j, \textbf{l})} = \spder{\phi}{r} \Bigr | _ {r(\bx_i, \bx_j, \textbf{l})} \pder{r}{r_c^k} \Bigr | _ {(\bx_i, \bx_j, \textbf{l})} \pder{r}{r_p^l} \Bigr | _ {(\bx_i, \bx_j, \textbf{l})} + \pder{\phi}{r} \Bigr | _ {r(\bx_i, \bx_j, \textbf{l})} \dfrac{\partial^2 r}{\partial r_c^k \partial r_p^l}  \Bigr | _ {(\bx_i, \bx_j, \textbf{l})}.
\end{align}
The first derivative of the kernel matrix $\bm{\mathsf{\Phi}}_d$ has a dimension of $nd \times n$ while the second derivative of the kernel matrix $\bm{\mathsf{\Phi}}_{dd}$ has a dimension of $nd \times nd$. They can be constructed according to
\begin{align}
    \bm{\mathsf{\Phi}}_d &= 
    \begin{pmatrix}
        \mathsf{\Phi}_{d_{11,1}} & \dots & \mathsf{\Phi}_{d_{11,d}} & \dots & \mathsf{\Phi}_{d_{1n,d}} \\
        \vdots & \ddots & \vdots & \ddots & \vdots \\
        \mathsf{\Phi}_{d_{n1,1}} & \dots & \mathsf{\Phi}_{d_{n1,d}} & \dots & \mathsf{\Phi}_{d_{nn,d}}
    \end{pmatrix}, \\
    \bm{\mathsf{\Phi}}_{dd} &= 
    \begin{pmatrix}
        \mathsf{\Phi}_{dd_{11,11}} & \dots & \mathsf{\Phi}_{dd_{11,1d}} & \dots & \mathsf{\Phi}_{dd_{1n,1d}} \\
        \vdots & \ddots & \vdots & \ddots & \vdots \\
        \mathsf{\Phi}_{dd_{11,d1}} & \dots & \mathsf{\Phi}_{dd_{11,dd}} & \dots & \mathsf{\Phi}_{dd_{1n,dd}} \\
        \vdots & \ddots & \vdots & \ddots & \vdots \\
        \mathsf{\Phi}_{dd_{n1,d1}} & \dots & \mathsf{\Phi}_{dd_{n1,dd}} & \dots &  \mathsf{\Phi}_{dd_{nn,dd}}
    \end{pmatrix}.
\end{align}

\subsubsection{Comparison of the interpolants constructed using RBF and GRBF}
In this section, we provide a comparison between RBF and GRBF models. We consider the Rastrigin function, given by $f(\bx) = 10d + \sum_{i=1}^d [x_i^2-10 \cos(2\pi x_i)]$, where $d$ is the number of dimensions of the input vector. To motivate the optimization of the internal parameters, surrogate models with different choices of internal parameters are considered to highlight their effect in the approximation accuracy.

\begin{figure}[htb]
    \centering
    \sidesubfloat[]{\includegraphics[width=0.45\textwidth]{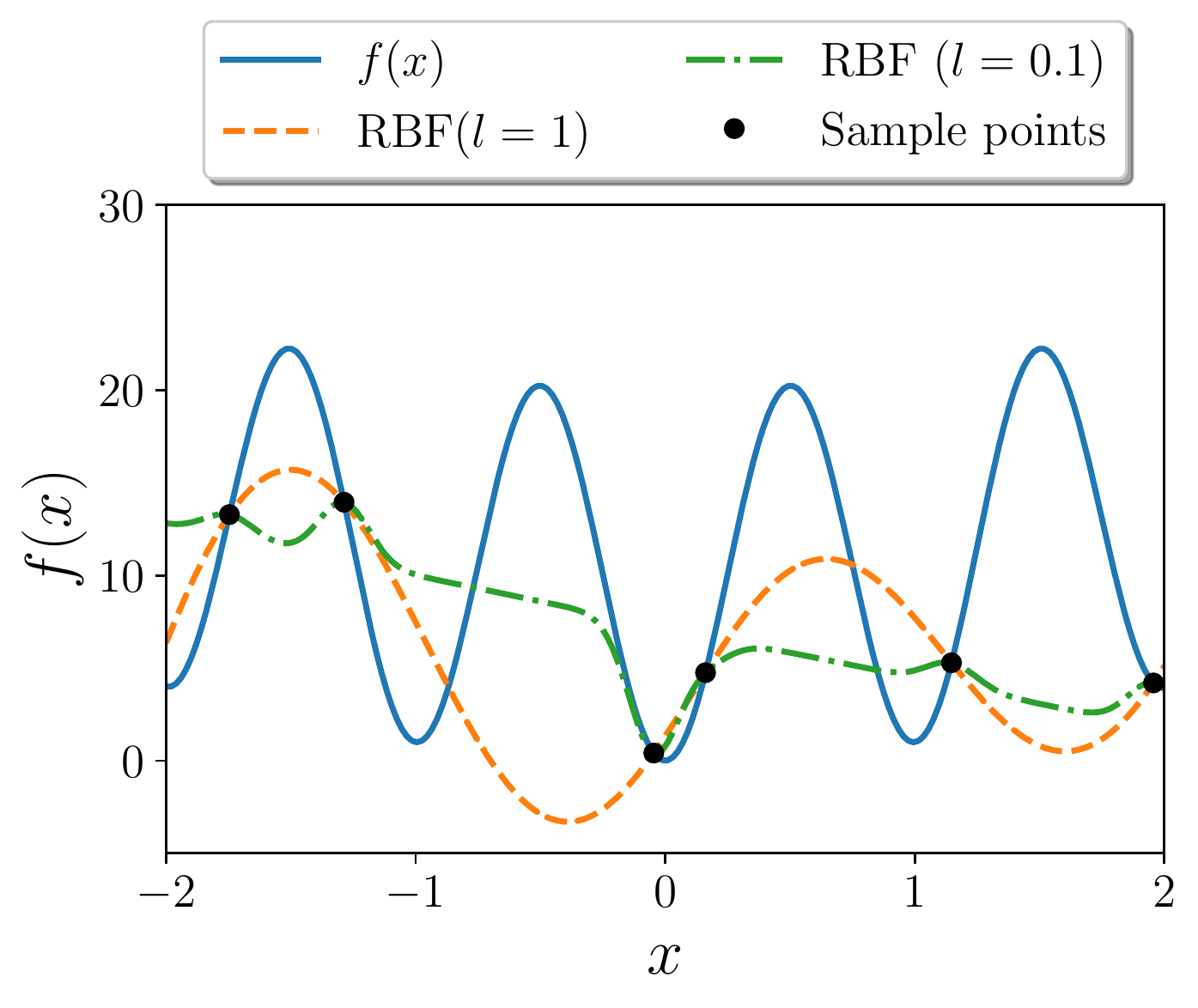} }
    \sidesubfloat[]{\includegraphics[width=0.45\textwidth]{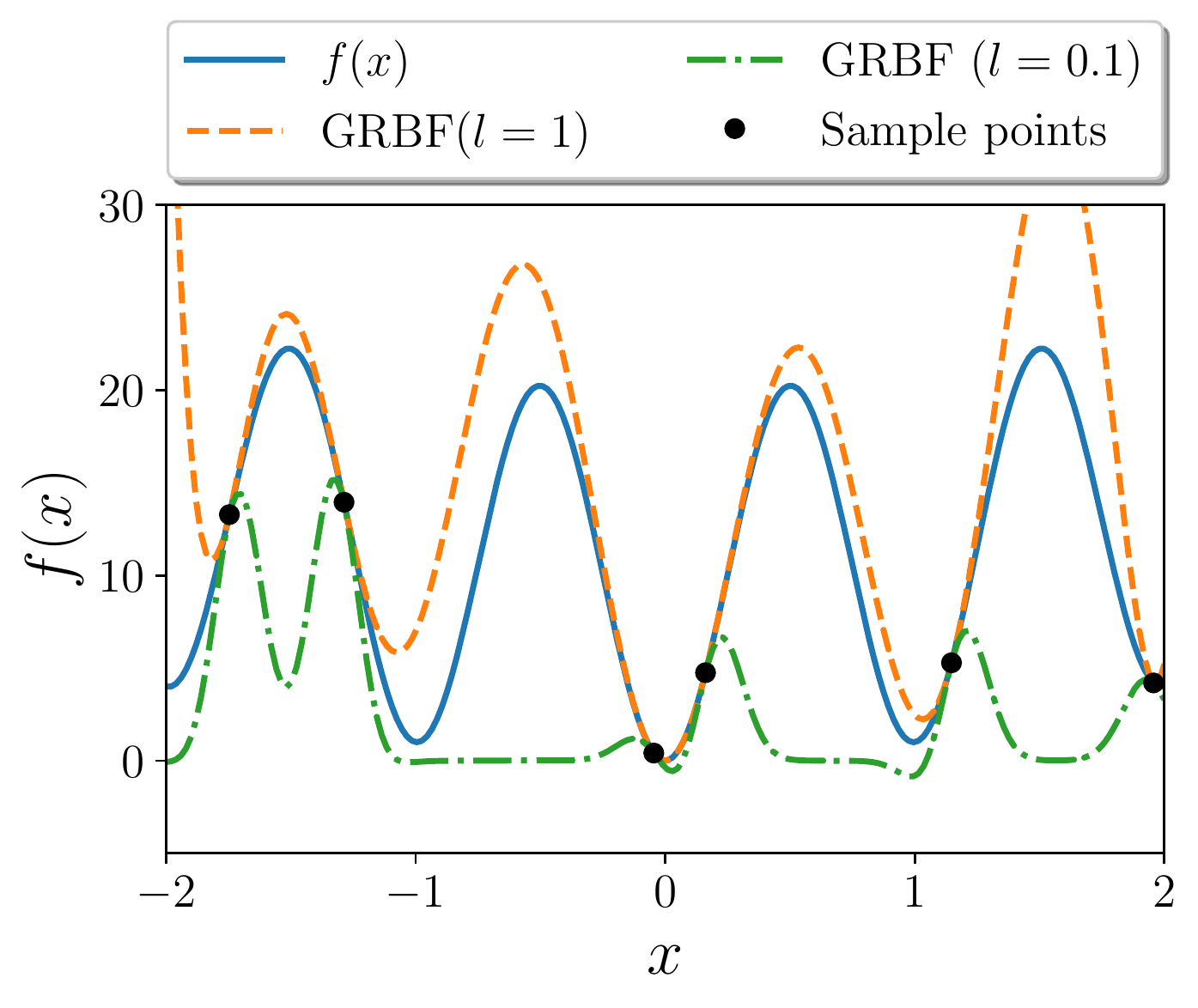} }
    \caption{Interpolation of the one-dimensional Rastrigin function (blue solid line) using 6 sample points (black points) (a) RBF with an exponential kernel and internal parameters $l=1.0$ (orange dashed line) and $l=0.1$ (green dash dotted line), (b) GRBF with an exponential kernel and internal parameters $l=1.0$ (orange dashed line) and $l=0.1$ (green dash dotted line).}
    \label{fig:rbf_1d}
\end{figure}

In the one-dimensional case, Fig.\ \ref{fig:rbf_1d}, interpolants based on the exponential kernel at six sample points and two different values of the internal parameter, $l=1.0$ and $l=0.1$, are considered. Fig.\ \ref{fig:rbf_1d}(a) presents the results using RBF whereas Fig.\ \ref{fig:rbf_1d}(b) shows the results using GRBF. This figure illustrates the effect of including gradient information in the surrogate model for different values of the internal parameter $l$. The interpolation achieved using GRBF with $l=1$ shows a considerable improvement with respect to the one given by RBF with the same value of $l$. Moreover, in both cases a large difference can be observed between the interpolation obtained with $l=1$ and that with $l=0.1$. These results suggest that the accurate construction of the surrogate using GRBF is highly dependent on the value of the internal parameter, otherwise adding the gradient information does not lead to a considerable improvement of the resulting interpolant. 

\subsubsection{Optimization of the internal parameters}
\label{sec:internal_params}
As shown in the previous section, the value of the internal parameter of the kernel function must be properly set to reach optimal performance in RBF and GRBF surrogate models. This can be achieved by optimizing the leave-one-out error $e_{loo}$ in the case of RBF as described in \cite{rippa_algorithm_1999}. In the case of GRBF, a different approach must be followed as we will show in this section. In this work, we make use of an efficient implementation of the leave-one-out error from \cite{bompard_surrogate_2010}, where the internal parameter is determined by the solution of the following optimization problem,
\begin{equation} \label{eq:e_loo}
    \min e_\mathrm{loo}(\mathbf{l}, \nu)\text,\qquad e_\mathrm{loo}(\mathbf{l}, \nu) = \dfrac{\mathbf{a}^T \bm{\mathsf{H}}(\mathbf{l}, \nu)^{-2} \mathbf{a}}{n \ \mathrm{diag}(\bm{\mathsf{H}}(\mathbf{l}, \nu)^{-2})} \qquad\mathrm{s.t.}\qquad\ \kappa(\bm{\mathsf{H}})<\dfrac{1}{10\epsilon},
\end{equation}
where $e_\mathrm{loo}$ is the leave-one-out error, $n$ is the number of evaluated points, vector $\mathbf{a}$ contains the values of the function at the evaluated points in the case of RBF and the values of the function and its gradient in the case of GRBF, $\kappa$ is the condition number of a matrix, $\epsilon$ is the machine precision, and $\bm{\mathsf{H}}$ can be defined as
\begin{equation} \label{eq:full_kernel}
    \bm{\mathsf{H}} = \bm{\mathsf{\Phi}},
\end{equation}
in the case of RBF, and as
\begin{equation} \label{eq:full_kernel_grbf}
    \bm{\mathsf{H}} = \begin{pmatrix}
        \bm{\mathsf{\Phi}} & \bm{\mathsf{\Phi}}_d \\ 
        -\bm{\mathsf{\Phi}}_d^T & \bm{\mathsf{\Phi}}_{dd}
    \end{pmatrix},
\end{equation}
in the case of GRBF. The constraint on the condition number of the full kernel matrix $\bm{\mathsf{H}}$ has been added to the optimization to ensure the smoothness of the surrogate model. The maximum value for the condition number is set to $1/10\epsilon$, where $\epsilon$ is the machine precision.

\begin{figure}[ht]
    \centering
    \sidesubfloat[]{\includegraphics[width=0.75\textwidth]{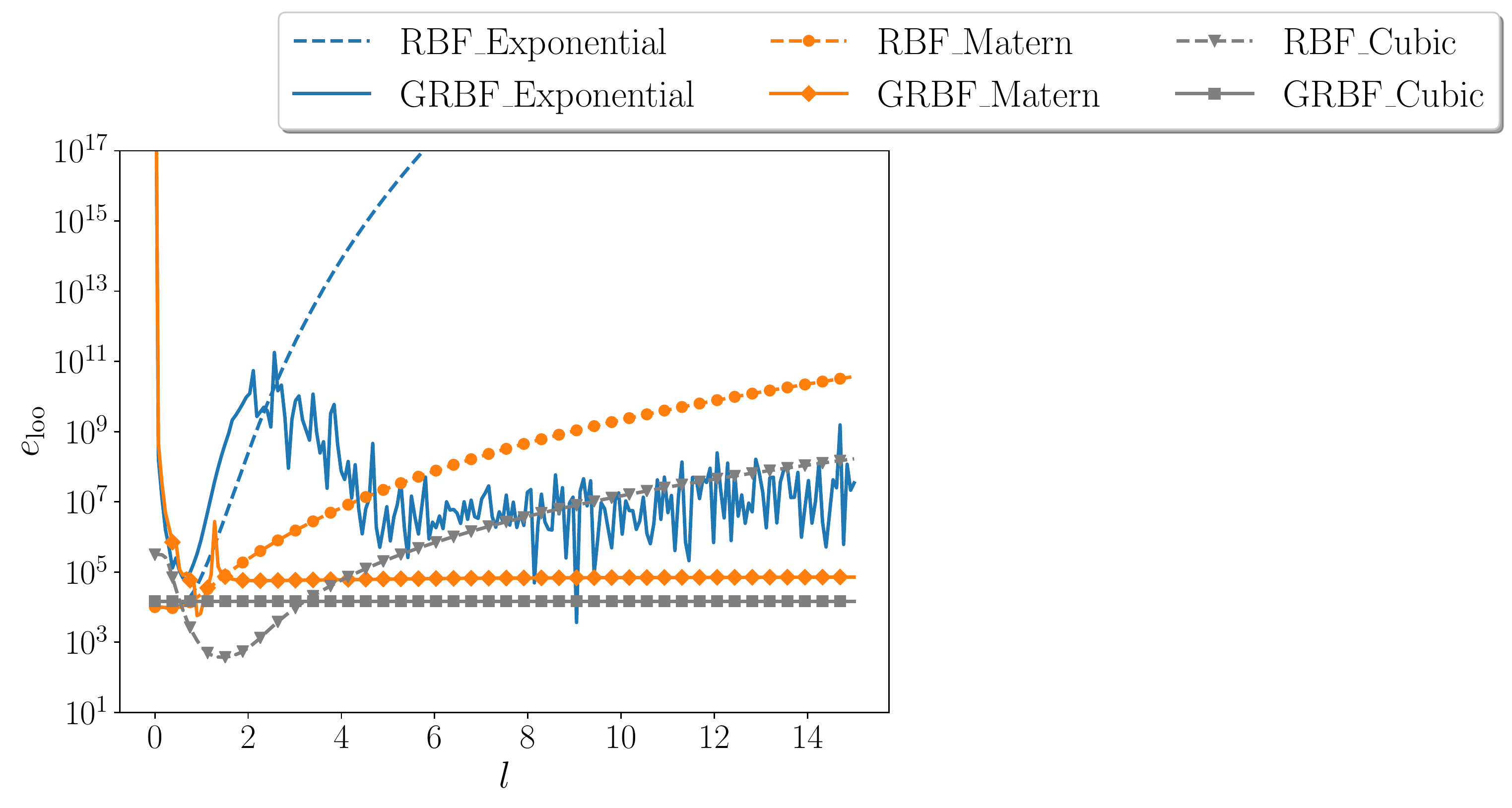} }
    \sidesubfloat[]{\hspace{-5cm}\includegraphics[width=0.45\textwidth]{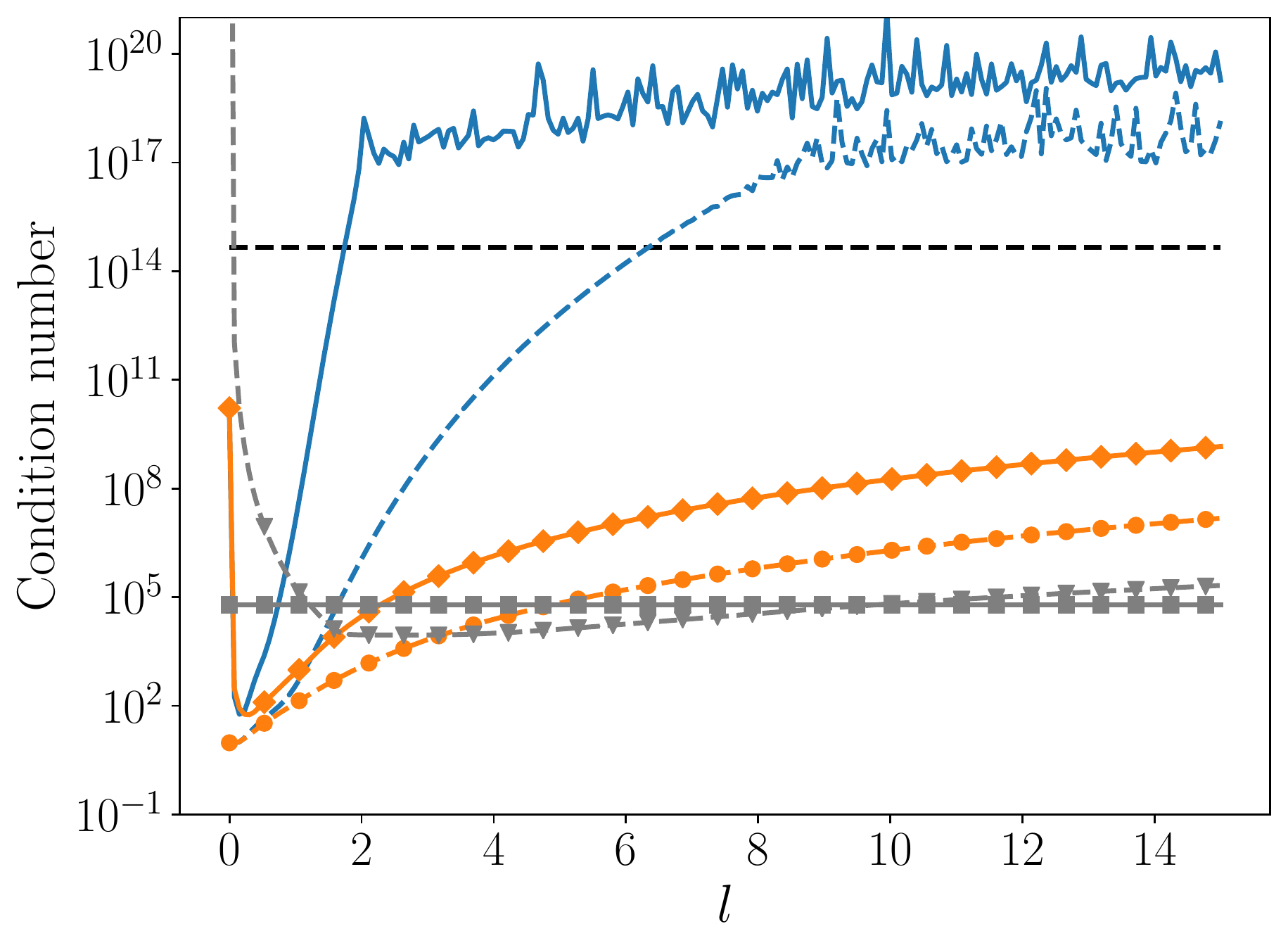} } \\
    \sidesubfloat[]{\includegraphics[width=0.45\textwidth]{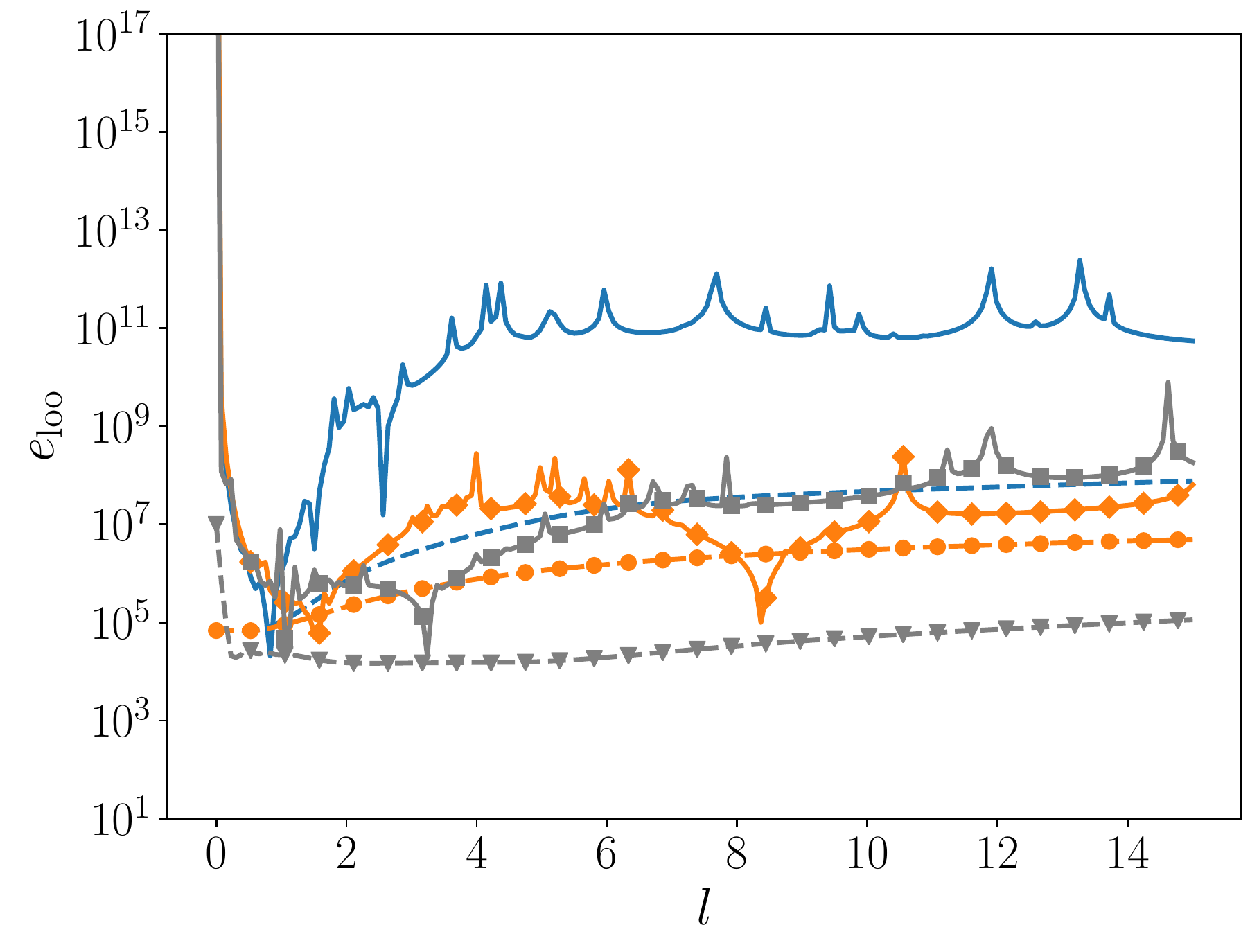} }
    \sidesubfloat[]{\includegraphics[width=0.45\textwidth]{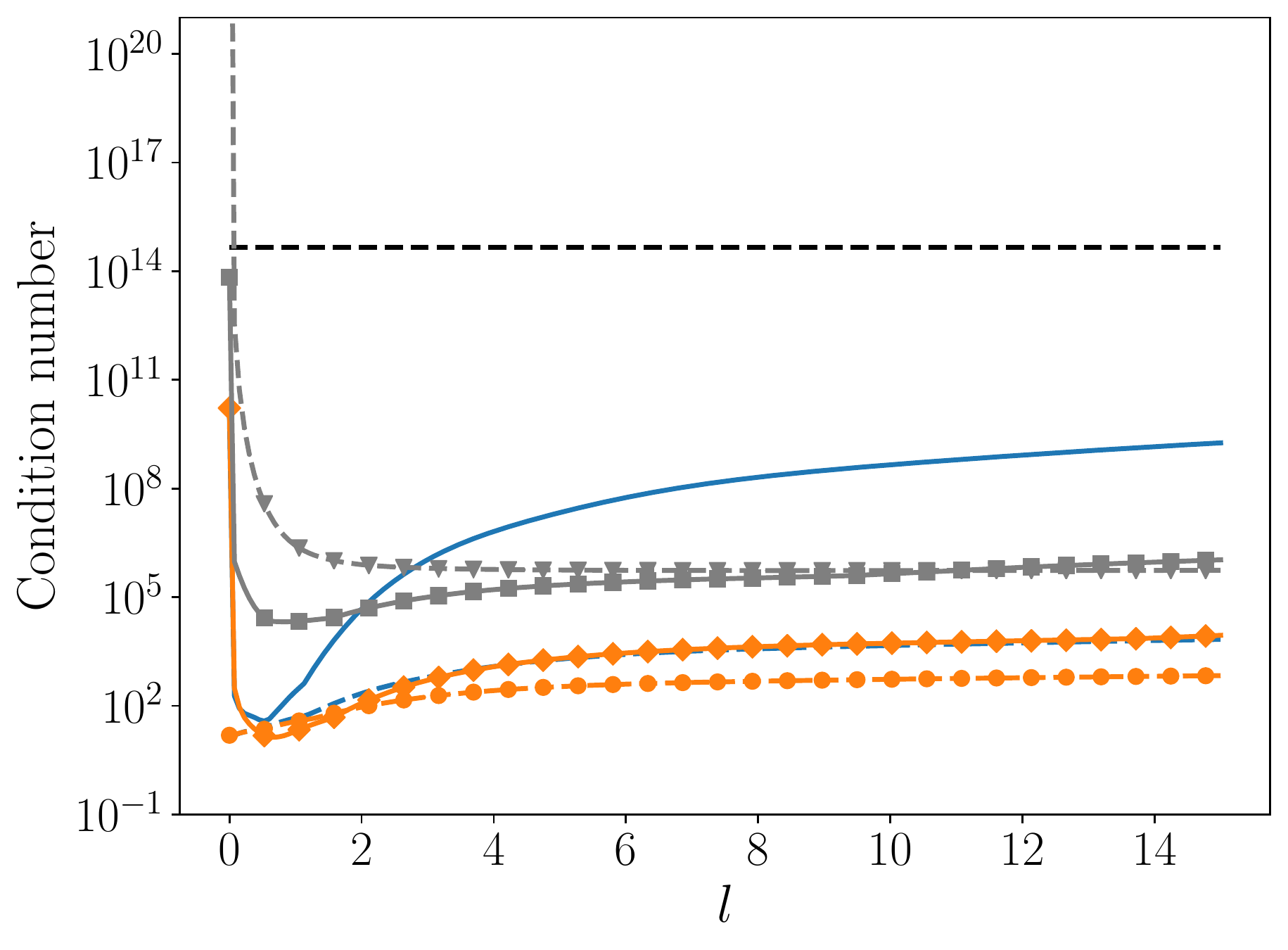} }
    \caption{Leave-one-out error as function of the internal parameter for (a) one-dimensional Rastrigin function and (c) two-dimensional Rastrigin function and condition number of the full kernel matrix for (b) one-dimensional Rastrigin function and (d) two-dimensional Rastrigin function. Results for the exponential kernel are plotted in blue (dashed line RBF and solid line GRFB), for the Matérn kernel are plotted in orange (dashed line with circles RBF and solid line with diamonds GRBF) and for the cubic kernel are plotted in grey (dashed line with triangles RBF and solid line with squares GRBF). The dashed black line represents the constraint on the condition number with a value of $1/(10\epsilon)$.}
    \label{fig:loo}
\end{figure}

Fig.\ \ref{fig:loo} displays the leave-one-out error and the condition number as a function of the internal parameter for two different cases and three kernel functions. Figs.\ \ref{fig:loo}(a,b) show the results for the one-dimensional Rastrigin function evaluated at 10 points and Figs.\ \ref{fig:loo}(c,d) present the results for the two-dimensional Rastrigin function evaluated at 20 points. In the latter, the internal parameter is kept constant in one direction and varies in the other. As it can be seen, the leave-one-out error presents a smooth behaviour for the RBF kernels when the constraint is satisfied, however when applied to GRBF kernels, the figure shows several peaks even though the condition number is below the constraint. In view of this, an optimal value for the internal parameter cannot be obtained through the optimization of the leave-one-out error in the case of GRBF surrogates. To circumvent this limitation, we propose instead to set the internal parameter $l$ to the inverse of the average absolute value of the derivatives in each direction obtained during the previous iterations of the optimization procedure when a gradient-enhanced kernel is employed. Directions with steeper derivatives are expected to feature smaller spatial scales, and therefore, the widths of the kernel can be reduced accordingly to approximate the objective function more accurately. Eq.\ \eqref{eq:optim_l} gives the expression used to compute the value of the internal parameter in this case,
\begin{equation} \label{eq:optim_l}
    \mathbf{l} = \left \lbrace \dfrac{1}{[ | \overline{\mathbf{g}|}(x ^ 1) | , \dots, | \overline{\mathbf{g}}(x^{d}) | ]} \right \rbrace.
\end{equation}

\subsection{Stochastic search algorithm}
\label{sec:algo}
In this section, we provide a description of the steps that are carried out to perform an optimization using the DYCORS algorithm.

DYCORS is a derivative-free stochastic optimization algorithm adapted to the optimization of high-dimensional expensive black-box functions. It was developed as a modification of the Local Metric Stochastic Response Surface (LMSRS) method \cite{regis_stochastic_2007} by introducing ideas from the Dynamically Dimensioned Search (DDS) method \cite{tolson_dynamically_2007}. In its original form, the algorithm does not rely on the gradient of the objective function to reach a minimum and therefore has no information on the shape of the objective function apart from its value for a given set of control parameters. 
The algorithm is detailed in Algorithm \ref{algo:main} as well as Algorithms \ref{algo:dycors}-\ref{algo:update} given in \ref{sec:app_algo}. The main steps of the algorithm are described below:\\

\noindent\textbf{1-Initialization}: The algorithm performs a fixed number of function evaluations $N_\mathrm{max}$. It is initialized by evaluating the objective function $f$ defined on the hypercube $\mathcal{D} = [a_d,b_d]\subseteq \mathbb{R}^d$ at a number of $m$ given initial sampling points $\mathcal{I}$. The initial sampling points can be generated by means of Latin Hypercube Sampling techniques. This method creates an optimal distribution through the full hypercube \cite{helton_latin_2003}. In this study, an enhanced Latin Hypercube Sampling based on \cite{beachkofski_improved_2002} is used to generate the initial sampling points $\mathcal{I}$. This method ensures that the minimum distance between the points is $d_{opt} = m/\sqrt[d]{m}$, and that each region of the hypercube has an equal representation on $\mathcal{I}$.
    
\noindent\textbf{2-Construction of the surrogate model}: At every iteration, a surrogate model is built following the procedure discussed in Section \ref{sec:surrogate}. The coefficients of the interpolant $\boldsymbol{\lambda}$ and $\textbf{c}$ are given by the solution of the linear systems in Eq.\ \eqref{eq:RBF_system} (RBF case) or \eqref{eq:GRBF} (GRBF case). 
    
\noindent\textbf{3-Generation  of trial points and evaluation using the surrogate model}: Following Algorithm \ref{algo:dycors}, the trial points are generated by perturbing the location of the evaluated point with the minimum function value in randomly selected directions. As the optimization procedure advances, the probability of perturbing a direction is reduced according to
    \begin{equation} \label{eq:phi}
        \varphi(n) = \varphi_0 \left(1 - \dfrac{\ln{(n-m+1)}}{\ln{(N_\mathrm{max}-m)}} \right),    
    \end{equation}
where $n$ is the number of function evaluations that have already been performed, $m$ is the size of the initial set of points, $N_{max}$ is the total number of function evaluations to be performed and $\varphi_0$ is a constant that will be defined later. Once the perturbed coordinates have been selected, $k$ trial points are generated by means of a normal distribution centered at the current minimum valued point with standard deviation $\sigma_n$. Due to the low computational cost of evaluating the trial points using the surrogate model, thousands of evaluations can be performed at a negligible cost. The value of the standard deviation varies depending on the number of consecutive failed or successful iterations, where a failed iteration means that the minimum valued point has not changed in the last iteration and a successful iteration means the algorithm has been able to improve the minimum. The initial value of the standard deviation is set to $0.2$ times the distance between boundaries of the hypercube in every direction. If $\tau_f$ consecutive failed iterations are performed, the standard deviation is divided by $2$. In case $\tau_s$ consecutive successful iterations are carried out, the standard deviation is multiplied by $2$. If the standard deviation falls below a given threshold $\sigma_m$, the algorithm is completely reinitialized to escape from local minima, by keeping just the information of the best evaluated point so far. Once the trial points have been generated, the surrogate model is evaluated at these points using Eq.\ \eqref{eq:interp_RBF}.
    
\noindent\textbf{4-Selection of best candidate point among the trial points}: In order to select the next point that will be evaluated using function $f$, we have to apply a selection criteria to the trial points. Algorithm \ref{algo:snp} provides the steps that are required to select this point. Using this selection criteria two different scores are given to each trial point. On the one hand, the first score (RBF score) takes into account the value of the surrogate model at the trial points, where the lowest value will get the best score. On the other hand, the second score (distance score) takes into account the distance between each trial point and all the already evaluated points, where the higher distances get better scores. The two scores are summed and the trial point with the best overall score is chosen as next point to be evaluated. Depending on the number of the current iteration, one of the scores may be given a greater weight in the overall score. The weight for the first score is rolled through the values $\boldsymbol{\Upsilon} = \left\lbrace \Upsilon_1, \Upsilon_2, \Upsilon_3, \Upsilon_4 \right\rbrace$ whereas the weights for the second score are one minus the value of the first score. By employing this scores, we ensure that different regions of the hypercube are populated, a mandatory criterion to avoid problems with singular matrices when building the surrogate model. This way of proceeding also helps to escape from local minima.
    
\noindent\textbf{5-Evaluation of the objective function at the best candidate point}: After selecting the best candidate point, the objective function (and its gradient in the gradient-enhanced case) is evaluated using the CFD solver. This is the most expensive step in the whole procedure as it requires to perform a full CFD simulation.
    
\noindent\textbf{6-Update information}: After evaluating the objective function, depending on the value obtained after, the counters that keep track of the consecutive failed and successful iterations can either be increased by one or set to zero, $C_\mathrm{f}$ and $C_\mathrm{s}$ respectively. If they reach the values $\tau_f$ or $\tau_s$, respectively, the value of the standard deviation used to generate the trial points $\sigma_n$ is modified accordingly. Afterwards, the set of evaluated points $\mathcal{A}_n$ and the iteration number $n$ are updated. These steps are indicated in Algorithm \ref{algo:update}.
    
\noindent\textbf{7-Optimization of the internal parameters}: Following Section \ref{sec:internal_params}, the internal parameters of the kernel function are optimized to improve the accuracy of the surrogate model. Every $n_\mathrm{ip}$ iterations of the algorithm, a differential evolution optimization algorithm is employed to optimize the values according to the leave-one-error \cite{rippa_algorithm_1999}. This step was not present in the original DYCORS algorithm. \\

Table \ref{tab:dycors_const} presents a summary of all the parameters used in the DYCORS algorithm, defined in \cite{regis_combining_2013}. The number of initial points $m$ is fixed to $m=d+1$ to ensure that singular matrices do not appear when building the RBF, although a higher value may be employed. The value of $\varphi_0$ is set such that in the first iteration of low-dimensional optimization problems $(d<20)$ all the coordinates are perturbed, whereas for higher dimensional problems, on average $20$ coordinates are perturbed at a time. The justification for this value of $\varphi_0$ is that the probability of improving the solution is increased if only a small amount of the variables are perturbed even at the beginning of the optimization procedure. The minimum standard deviation $\sigma_m$ allows the reduction of the standard deviation up to $6$ times before the algorithm is restarted to ensure that local minima are skipped. The weight pattern $\boldsymbol{\Upsilon}$ starts with a value that gives more importance to the distance score and progressively increases the importance of the RBF score in the overall score.

\begin{table}
    \centering
    \begin{tabular}{l l c}
        Parameter & Description & Value \\
        \hline
        $m$ & Number of initial sampling points & $d+1$ \\
        $k$ & Number of trial points to be generated & $\min (100d,5000)$ \\
        $\varphi_0$ & Initial probability of perturbing a direction & $\min(20/d,1)$ \\
        $\sigma_0$ & Initial standard deviation & $0.2 (b_d-a_d)$ \\
        $\sigma_m$ & Minimum standard deviation & $0.2/2^6 (b_d-a_d)$ \\
        $\tau_s$ & Maximum number of consecutive successful iterations & $3$ \\
        $\tau_f$ & Maximum number of consecutive failed iterations & $5$ \\
        $\boldsymbol{\Upsilon}$ & Weight pattern in the score of the trial points & $\left\lbrace 0.3, 0.5, 0.8, 0.95 \right\rbrace$ \\
    \end{tabular}
    \caption{DYCORS parameters.}
    \label{tab:dycors_const}
\end{table}

\begin{algorithm}
    \DontPrintSemicolon
    \SetAlgoLined
    \KwIn{Real valued black-box function, $f$ defined on $\mathcal{D} = [a_d,b_d]\subseteq \mathbb{R}^d$\newline
        Real valued black-box function, $g$ defined on $\mathcal{D} = [a_d,b_d]\subseteq \mathbb{R}^d$ in case of G-DYCORS\newline
        Maximum number of function evaluations, $N_\mathrm{max}$ \newline
        Initial and minimum standard deviations, $\sigma_0$ and $\sigma_m$ \newline
        Number of trial points, $k$ \newline
        Response surface model, $\phi$ \newline
        Interpolant, $s_n$ \newline
        Internal parameter of the kernel, $\mathbf{l}$ \newline
        Initial sampling points, $\mathcal{I}=\left\lbrace \bx_1,\dots,\bx_{m} \right\rbrace$ \newline
        Weight pattern, $\boldsymbol{\Upsilon} = \left\lbrace \Upsilon_0, \Upsilon_1, \Upsilon_2, \Upsilon_3 \right\rbrace$ \newline
        Limits for number of consecutive failed and successful iterations, $\tau_\mathrm{f}$ and $\tau_\mathrm{s}$ \newline
        Number of iterations without optimizing the internal parameter, $n_\mathrm{ip}$ \newline
    }
    \KwResult{Best point encountered, $x_\mathrm{best}$} \;
    
    \textit{\textbf{Initialize algorithm}}: $\mathcal{A}_m = \mathcal{I}, f(\bx), (g(\bx)) : \bx\in\mathcal{A}_m$\;
    
    \textit{\textbf{Select best evaluated point}}: $f_\mathrm{best} = f(\bx_\mathrm{best})$\;
    
    \textit{\textbf{Initialize standard deviation and counters}}: $\sigma_n = \sigma_0,\ n=m,\ C_\mathrm{f} = 0$ and $C_\mathrm{s} = 0$ \;
    
    \While{$n<N_\mathrm{max}$}{         
       \textit{\textbf{Construct the surrogate model}}: Compute $\boldsymbol{\lambda}$ and $\mathbf{c}$ following Sections \ref{sec:RBF} and \ref{sec:GRBF}\;
        
       \textit{\textbf{Generate and evaluate trial points}}: Algorithm \ref{algo:dycors}: trial\_points($n, k, \sigma_n, s_n, \mathcal{A}_n, \boldsymbol{\lambda}, \mathbf{c}$)\;
        
        \textit{\textbf{Select best candidate point}}: Algorithm \ref{algo:snp}: select\_next\_point($n, \boldsymbol{\Upsilon}, \mathcal{A}_n, \textbf{y}_{n,j}, s_n(\textbf{y}_{n,j})$)\;
        
        \textit{\textbf{Evaluate function (and gradient)}}: Compute $f(\bx_{n+1})$ , ($g(\bx_{n+1})$)\;
        
        \textit{\textbf{Update information}}: Algorithm \ref{algo:update}: update\_info($n, \bx_\mathrm{best}, f_\mathrm{best}, \bx_{n+1}, f_{n+1}, C_\mathrm{f}, C_\mathrm{s}, \tau_\mathrm{f}, \tau_\mathrm{s}, \sigma_n, \mathcal{A}_n$)
        
        \textit{\textbf{Optimize internal parameters}}: following Section \ref{sec:internal_params}\;
    }
    \caption{(G)-DYCORS algorithm}
    \label{algo:main}
\end{algorithm}

\section{Governing equations}
\label{sec:equations}
The flow solver employed in this study implements the projection-based immersed boundary method from \cite{taira_immersed_2007} for two-dimensional flows. The governing equations in continuous form
\begin{align}
    \pder{\bu}{t} + \bu \cdot \nabla \bu &= -\nabla p + \dfrac{1}{Re} \nabla^2 \bu + \int_\mathcal{C}\mathbf{f}(s, \mathbf{x}) \delta(\mathbf{\hat{x}}-\xi(s))\,\mathrm{d}s, \label{eq:momentum} \\
    \nabla \cdot \bu &= 0, \label{eq:divergence} \\
\intertext{and}
    \bu[\xi(s)] &= \int_\mathcal{M}\bu(\mathbf{\hat{x}})\delta(\xi(s)-\mathbf{\hat{x}})\,\mathrm{d}\mathbf{\hat{x}} = \bu_B(s), s \in \mathcal{C}, \label{eq:boundary}
\end{align}
are solved on a given domain $\mathcal{M}$ together with suitable initial and boundary conditions. In the above, $\mathbf{\hat{x}} \in \mathcal{M}, \mathbf{u}$, $p$, $\mathbf{f}(s,\bx), \bx$ and $Re$ are, respectively, the velocity vector, the pressure, the distributed momentum sources along the boundaries of the solids $\mathcal{C}$, the set of control parameters that define the boundary force when an actuation wants to be applied on the surface, and the Reynolds number. The pressure $p$ and the boundary force $\mathbf{f}(s,\bx)$ can be regarded as a set of Lagrange multipliers that enforce the incompressibility constraint and the no-slip boundary condition or the actuation on $\mathcal{C}$, respectively. A staggered-mesh finite-volume formulation is used to discretize Eqs.\ \eqref{eq:momentum}-\eqref{eq:boundary} using the implicit Crank-Nicolson integration method for the viscous terms and the explicit second-order Adams-Bashforth scheme for the advection terms. The integrals that involve the $\delta$ function are discretized using the mollified $\delta$ function from \cite{roma_adaptive_1999}. The resulting discretized governing equations then are
\begin{equation} \label{eq:discretized-eqs}
    \begin{pmatrix}
    \bm{\mathsf{A}} & \bm{\mathsf{Q}} \\ 
    \bm{\mathsf{Q}}^T & \bm{\mathsf{0}}
    \end{pmatrix} \begin{pmatrix}
    \mathbf{q}^{k+1} \\ 
    \boldsymbol{\lambda}
    \end{pmatrix} = \begin{pmatrix}
    \bm{\mathsf{B}}\mathbf{q}^k -\frac{3}{2} \mathcal{N}(\mathbf{q}^k) + \frac{1}{2} \mathcal{N}(\mathbf{q}^{k-1}) + \mathbf{bc}_1 \\ 
    \mathbf{r}_2
    \end{pmatrix},
\end{equation}
or in compact form
\begin{equation}\label{eq:discretized-eqs-compact}
R(\mathbf{q}^{k-2}, \mathbf{q}^{k-1}, \mathbf{q}^k, \mathbf{x}) = 0.
\end{equation}
In the above, $\mathbf{q}^{k}$ and $\boldsymbol{\lambda}$ are the flow field at a given time step and the Lagrange multipliers. The reader is referred to \cite{taira_immersed_2007} for further details regarding the various definitions of the matrices $\bm{\mathsf{A}},\ \bm{\mathsf{Q}}$ and $\bm{\mathsf{B}}$, the nonlinear function $\mathcal{N(\cdot)}$ and the vectors $\mathbf{bc}_1$ and $\mathbf{r}_2$. The solver is equipped with the linearized direct and adjoint equations respectively (see \ref{sec:app_adjoint}), that allow the computation of the gradients using adjoint-based methods. The numerical solver IBMOS (Immersed Boundary Method for Optimization and Stability analysis) is available at \cite{fosas_de_pando_ibmos_2020}.

\section{Results}
\label{sec:results}
In this section, we first provide a description of the test cases that are employed in the optimization problem. Afterwards, the objective function and the control parameters are presented. Finally, the results given by the different optimization algorithms are discussed and compared.

\subsection{Problem description}
\begin{figure}[ht]
    \centering
    \includegraphics[width=\linewidth]{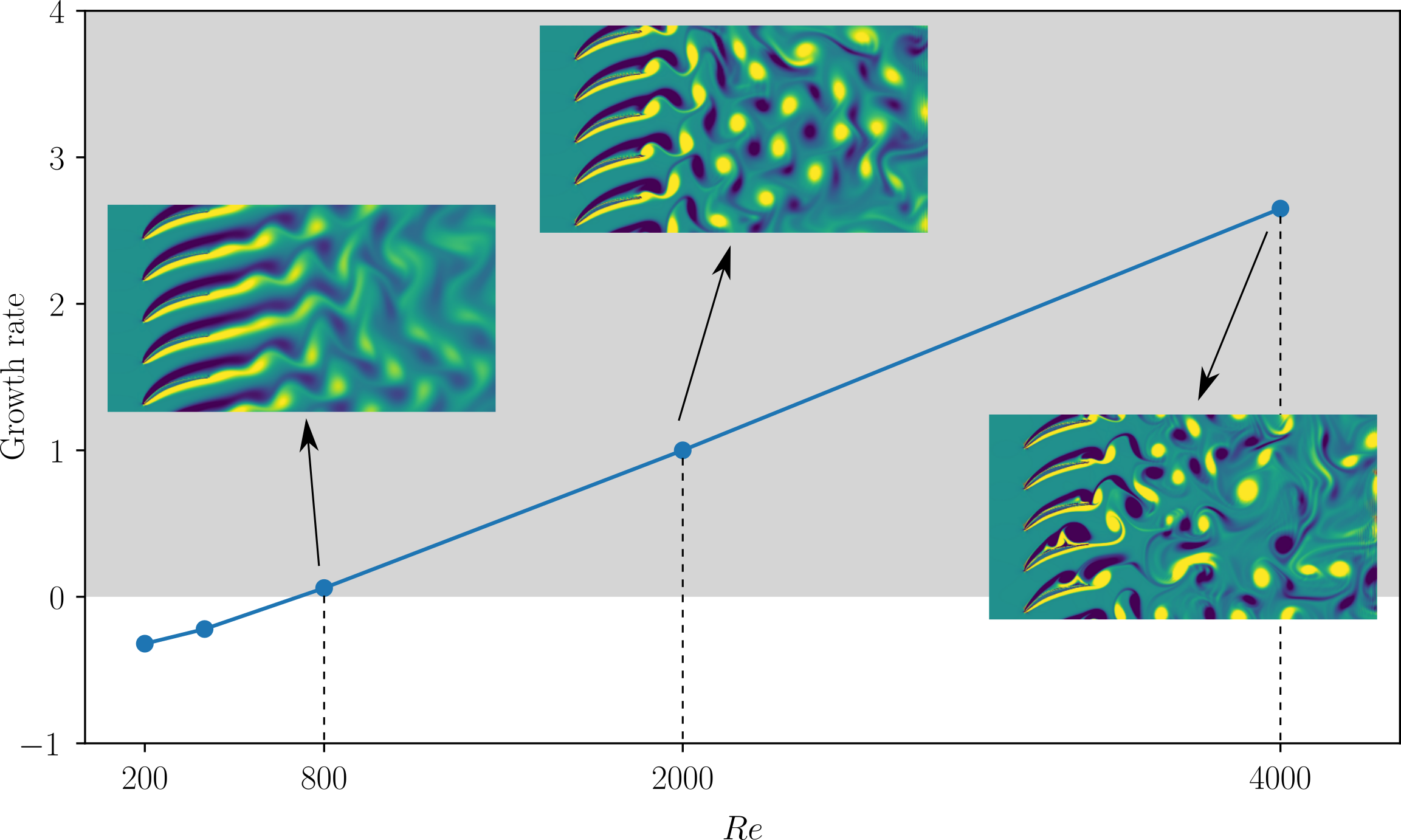}
    \caption{Growth rates of the leading modes at different $Re$ numbers and instantaneous snapshot showing the vorticity levels at the selected $Re$ numbers.}
    \label{fig:wi_re}
\end{figure}

The flow around a linear cascade consisting of five blades is used to assess the effectiveness of the stochastic optimization algorithm described in Section~\ref{sec:optimization_framework}. The chosen blade profile was developed in \cite{sanger_use_1983} and its aerodynamic characteristics have been extensively investigated experimentally and numerically~\cite{yang_experimental_2004, he_two-scale_2017, phan_validation_2020}. In the following, the stagger angle of the blades is set to $22.5^\circ$ and the angle of attack is $32.5^\circ$. Periodic boundary conditions are specified along the vertical direction, the velocity components are imposed at the inlet, and a convective outflow boundary condition is used at the outlet boundary. A representative snapshot of this flow at variable Reynolds numbers, depicted by instantaneous levels of vorticity~$\omega_z$, is shown in Fig.\ \ref{fig:wi_re}. 

A linear stability analysis has been performed to determine the critical Reynolds number. The growth rate of the leading mode for varying~$Re$ is shown in Fig.\ \ref{fig:wi_re}, suggesting that the critical Reynolds number for this configuration is $Re_c \approx 750$. To assess the efficiency of the optimization algorithm, representative examples around and far from criticality have been chosen at, respectively, $Re = \left\lbrace 800, 2000, 4000 \right\rbrace$, shown in Fig.\ \ref{fig:wi_re}. At $Re=800$, which is slightly above the critical Reynolds number, the flow presents an instability developing in the wake of the blades. As the Reynolds number is increased up to $Re=2000$, an instability develops upstream resulting in pairs of vortices shedding from the trailing edges of the blades. In this case, stronger interaction between the wakes of the different blades is observed, although the wake still displays a regular pattern. Finally, at $Re=4000$, the figure shows vortex shedding from the suction side close to the leading edge. Vorticity levels are higher in this case in comparison with the previous Reynolds numbers and a stronger interaction between the wakes is displayed, which leads to a chaotic behaviour downstream.

Table \ref{tab:test_cases} gives details on the numerical grids that have been used at each Reynolds number, consisting of a structured rectangular mesh stretched in the horizontal direction in the region around the blades. The vertical grid spacing remains uniform across the full computational domain. Both $Re=\{ 800,2000 \}$ use the same grid. Numerical grids with larger domain size and finer grid spacing were considered at these Reynolds numbers but no significant differences were observed neither in the spectrum nor in the spatial structures of the modes obtained with the stability analysis and therefore the flow is considered to be well resolved. At $Re = 4000$, a refined grid was considered to avoid numerical instabilities.

\begin{table}[ht]
    \centering
    \begin{tabular}{c c c c c}
        Reynolds & $[x_\mathrm{min},x_\mathrm{max}] \times [y_\mathrm{min},y_\mathrm{max}]$ & $\Delta x_\mathrm{min}$ & $\Delta x_\mathrm{max}$ & $\Delta y$ \\
        \hline
        $800$ & $[-4.43, 4.62] \times [-0.3, 0.3]$ & $0.006$ & $0.04$ & $0.006$ \\
        $2000$ & $[-4.43, 4.62] \times [-0.3, 0.3]$ & $0.006$ & $0.04$ & $0.006$ \\
        $4000$ & $[-4.72, 4.69] \times [-0.3, 0.3]$ & $0.0045$ & $0.03$ & $0.0045$
    \end{tabular}
    \caption{Grid parameters.}
    \label{tab:test_cases}
\end{table}

\subsection{Objective function and actuation}
\begin{figure}[ht]
    \centering
    \sidesubfloat[]{\includegraphics[width=0.45\textwidth]{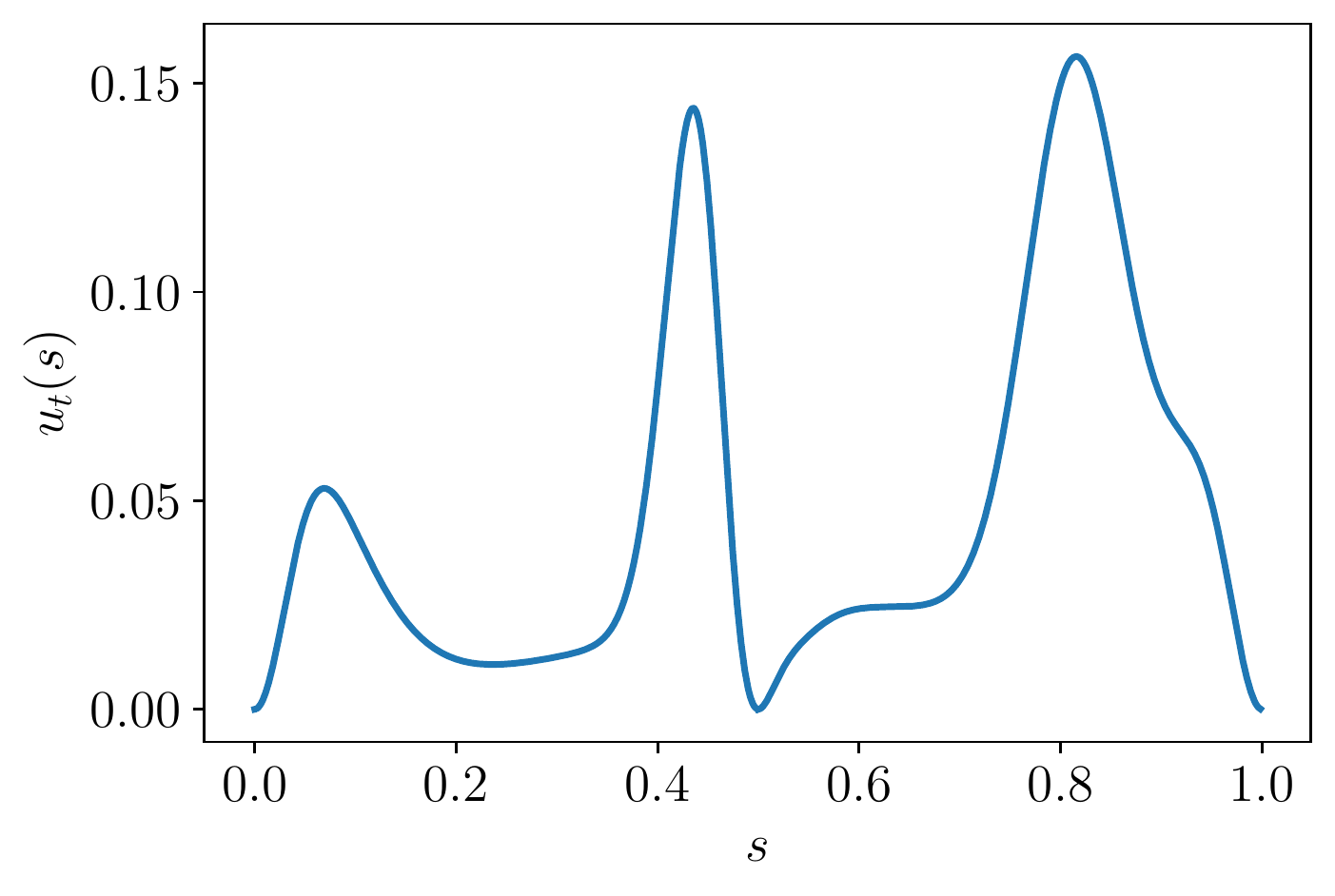} }
    \sidesubfloat[]{\includegraphics[width=0.45\textwidth]{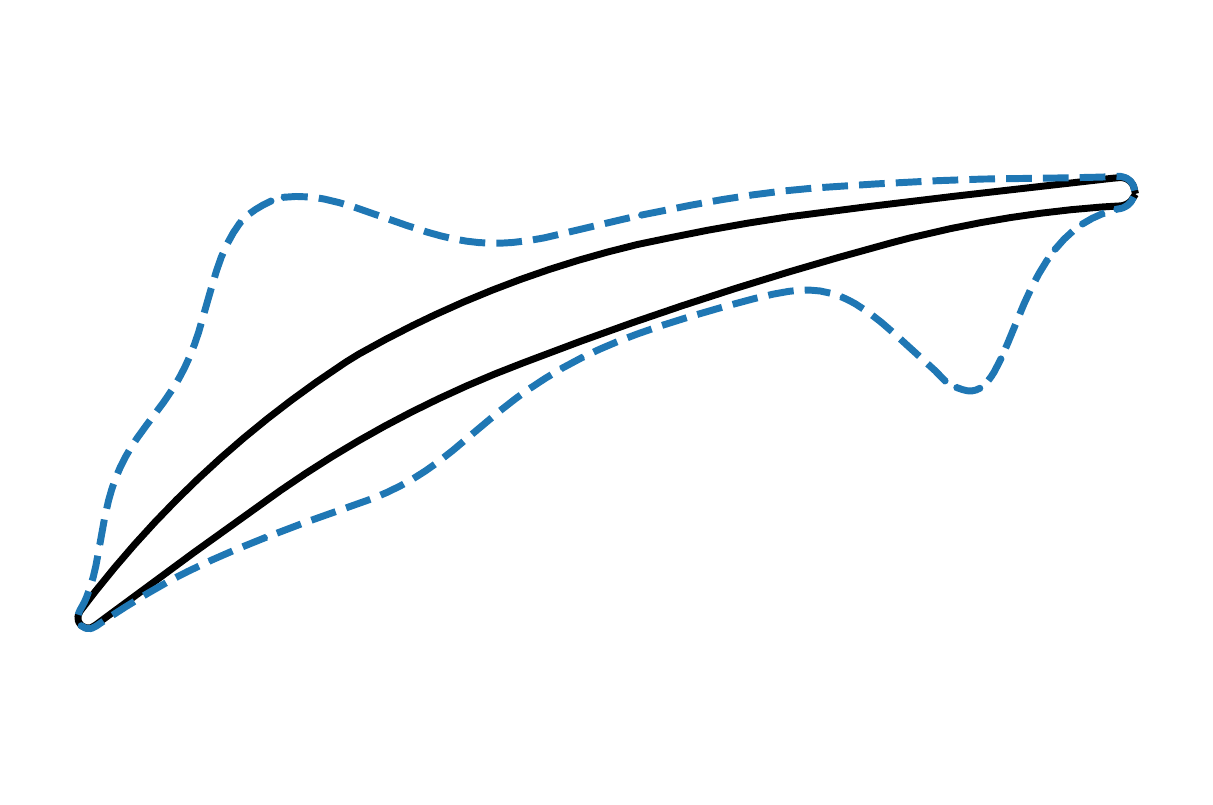} }
    \caption{Example of a representative actuation. (a) presents the tangential velocity as function of the arc-length $s$ while (b) shows the actuation (solid blue line) superposed to the blade profile (black dashed line).}
    \label{fig:actuation}
\end{figure}

We now intend to minimize the total pressure loss through the blade by means of an actuation that imposes a tangential velocity on the blade surface. The optimization problem can be stated as follows
\begin{equation}
    \min \mathcal{J}(\mathbf{q}^0,\dots,\mathbf{q}^K,\bx)\ \qquad\mathrm{s.t.}\qquad\  R(\mathbf{q}^{i-2}, \mathbf{q}^{i-1}, \mathbf{q}^{i}, \bx) = 0\ \forall\ i \in \{ 1,\dots,K \},
\end{equation}
where $\mathbf{q}^i$ is the state vector at the $i$-th time step, $\mathbf{q}^0$ is the initial condition (by convention, $\mathbf{q}^{-1} = \mathbf{q}^0$), $\bx$ is the set of control parameters, $R$ is the residual of the propagator that allows us to determine $\mathbf{q}^i$ as a function of $\mathbf{q}^{i-1}$ and $\mathbf{q}^{i-2}$, $K$ is the total number of iterations of the simulation and $\mathcal{J}$ is the objective function. The objective function is the defined by the sum of two terms: the average total pressure loss through the blade and a penalization term for the actuation. More precisely,
\begin{equation}
    \mathcal{J}(\mathbf{q}^0,\dots,\mathbf{q}^K,\bx) = \overline{\Delta p_0(\mathbf{q}^{K_0},\dots,\mathbf{q}^{K})} + \alpha \left \| u_t(\bx) \right \|,
\end{equation}
where $K_0$ is the index of the first time step that is considered in the temporal average of the total pressure loss and $\alpha$ is a positive constant that penalizes the strength of the actuation. Note that the parameter $K_0 > 1$ is set to remove the contribution of the initial transients from the cost function. The total number of iterations of the simulations $K$ is not fixed. Instead, it is updated dynamically at every simulation by applying the Cauchy criterion \cite{abbott_understanding_2015} to the averaged total pressure loss. The Hann windowing function \cite{mcbean_aspects_1993} is employed to speed up convergence. The Cauchy criterion ensures that every simulation has a large enough time window and consequently low frequencies are not bypassed. The prescribed tangential velocity on the blade surface is given by
\begin{equation} \label{eq:actuation}
    u_t(s, t) = \sum_{i=1}^n a_i f(2\pi s, 2\pi s_i, \sigma_i) \cos(\omega_i t + \phi_i),
\end{equation}
where $s$ is the position on the blade surface measured by the arc-length, $t$ is the time, $n$ is the number of actuators distributed over the surface, $a_i$ is the amplitude of the actuator, $s_i$ is the location of the maximum velocity imposed by the actuator, $\sigma_i$ sets the width of the actuator, $\omega_i$ is the frequency of the actuator and $\phi_i$ is the phase. The trailing edge corresponds to $s = 0.5$ whereas the leading edge corresponds to $s = 0$ on the pressure side and $s = 1$ on the suction side. Therefore, the pressure side corresponds to values of $s$ in the range $[0,0.5]$ and the suction side of the blade corresponds to values of $s$ in the range $[0.5,1]$. Details on function $f$ are given in \ref{sec:app_actuation}. The set of control parameters for blade $j$ is given by $\bx_j = (a_{1,j}, \dots, a_{n,j}, s_{1,j}, \dots, s_{n,j}, \sigma_{1,j}, \dots, \sigma_{n,j}, \omega_{1,j}, \dots, \omega_{n,j}, \phi_{1,j}, \dots, \phi_{n,j})$. An example of a representative actuation with four actuators is shown in Fig.\ \ref{fig:actuation}, where the maximum amplitude of each actuator, without taking into account the time-dependent term, is considered for the sake of clarity.

The flow around the blades is optimized by means of four actuators on each blade. The location of the actuators is constrained so that two actuators are located on each side. The maximum width of an actuator is fixed to half the arc-length of the blade profile, and the minimum to fifty times the minimum grid size to avoid steep gradients at the surface. The upper bound on the frequency parameters is set to four times the frequency of the leading mode at the corresponding $Re$ number whereas the lower bound is set to zero. The amplitude, location, width, and angular frequency of the actuators are taken the same for every blade, and a difference in phase $\varphi_j$ is allowed. More precisely, the phase of the $i$-th actuator on the $j$-th blade is given by $\phi_i + \varphi_j$, and by setting $\varphi_1 = 0$, the first blade is used as reference. The full set of 24 control parameters is then $\bx = (a_1, \dots, a_4, s_1, \dots, s_4, \sigma_1, \dots, \sigma_4, \omega_1, \dots, \omega_4, \phi_1, \dots, \phi_4, \varphi_2, \dots, \varphi_5)$.

\subsection{Performance of optimization strategies}
The effectiveness of the gradient-enhanced DYCORS algorithm is assessed by comparison against the original derivative-free version of DYCORS for simulations at $Re = \left\lbrace 800, 2000, 4000 \right\rbrace$. Cases with and without optimization of the internal parameters of the kernel are presented for $Re = \left\lbrace 800, 2000 \right\rbrace$. The gradient-based alternative L-BFGS-B \cite{byrd_limited_1995}, which uses a limited memory version of the BFGS algorithm \cite{fletcher_practical_1987} to approximate the Hessian matrix is also used at all Reynolds numbers to compare the stochastic-based algorithm with the gradient-based counterpart. All the surrogate model based optimizations for a given Reynolds number are initialized using the same initial sampling points and the gradient-based optimization is initialized using a random point from this initial sample. The optimizations performed using the derivative-free version of the DYCORS algorithm are limited to $N_\mathrm{max} = 250$ iterations, while the optimizations carried out using the gradient-enhanced version and the L-BFGS-B algorithm are limited to $N_\mathrm{max} = 125$ iterations. Therefore, all the optimizations employ the same CPU time as the cost of computing the gradient of the objective function, using our solver, is roughly the same as the cost of performing a single function evaluation.

The value of the objective function for the optimal set of control parameters, the values of the average total pressure drop, and the penalization term for each optimization case are given in Table \ref{tab:results}, where methods with the subscript $_{ip}$ indicate the cases with optimized internal parameters. According to this table, we can see that the gradient-enhanced version of DYCORS obtains the best results at $Re=2000$ and $Re=4000$ while the derivative-free version performs the best at $Re=800$. Moreover, updating the internal parameters of the kernel improves the solution in both versions of the algorithm, and as expected the GRBF surrogates do not perform satisfactorily when internal parameters are not optimized.

\begin{table}[ht]
    \centering
    \begin{tabular}{|l|c c c|c c c|c c c|}
        \hline
        \multicolumn{1}{|c|}{} & \multicolumn{3}{|c|}{$Re=800$} & \multicolumn{3}{|c|}{$Re=2000$} & \multicolumn{3}{|c|}{$Re=4000$} \\
        \hline
        \textbf{Methods} & \multicolumn{1}{l}{$f_{\mathrm{min}}$} & \multicolumn{1}{l}{$\overline{\Delta p_0}$} & \multicolumn{1}{l}{$\alpha \left \| u_t(\bx) \right \|$} & \multicolumn{1}{|l}{$f_{\mathrm{min}}$} & \multicolumn{1}{l}{$\overline{\Delta p_0}$} & \multicolumn{1}{l}{$\alpha \left \| u_t(\bx) \right \|$} & \multicolumn{1}{|l}{$f_{\mathrm{min}}$} & \multicolumn{1}{l}{$\overline{\Delta p_0}$} & \multicolumn{1}{l|}{$\alpha \left \| u_t(\bx) \right \|$} \\
        \hline
        \textbf{No actuation} & 0.2099 & 0.2099 & 0.0 & 0.1071 & 0.1071 & 0.0 & 0.1277 & 0.1277 & 0.0 \\
        \textbf{L-BFGS-B} & 0.1965 & 0.1882 & 0.0083 & 0.1084 & 0.1072 & 0.0012 & 0.1246 & 0.1245 & 0.0001 \\
        \textbf{DYCORS} & 0.1710 & 0.1648 & 0.0062 & 0.1069 & 0.1068 & 0.0001 & - & - & - \\
        \textbf{DYCORS$_{ip}$} & 0.1691 & 0.1625 & 0.0066 & 0.1059 & 0.1057 & 0.0002 & 0.1109 & 0.1105 & 0.0004 \\
        \textbf{G-DYCORS} & 0.1981 & 0.1955 & 0.0026 & 0.1060 & 0.1057 & 0.0003 & - & - & - \\
        \textbf{G-DYCORS$_{ip}$} & 0.1730 & 0.1657 & 0.0073 & 0.1039 & 0.1036 & 0.0003 & 0.1085 & 0.1083 & 0.0002 \\
        \hline
    \end{tabular}
    \caption{Optimization results.}
    \label{tab:results}
\end{table}

Considering the gradient-based algorithm L-BFGS-B, the table shows that it presents a performance comparable to that of the stochastic algorithms only at $Re=800$. At low Reynolds numbers, the gradient-based algorithm is expected to provide good results since the probability for the presence of multiple local minima is small due to the deterministic nature of the flow. This also implies that the different versions of the stochastic algorithms may not show significant differences, since the computed gradients may not be too steep, resulting in a comparable estimation of the interpolant using either derivative-free or gradient-enhanced version of the algorithm. 

At $Re=2000$, however, the L-BFGS-B algorithm is not even able to improve upon the case without actuation. This behaviour can be explained by the fact that the objective function is expected to have a larger amount of local minima due to the increase in the chaotic nature of flow as the Reynolds number increases, illustrated by comparing the vortical structures shown in Fig.\ \ref{fig:wi_re}. The presence of multiple local minima degrades the performance of gradient-based algorithms which are prone to get stuck in local minima. Also, presence of steeper gradients in the objective function means that derivative-free surrogates should not be able to properly interpolate the objective function and that adding the gradient information should improve the construction of the interpolant, resulting in the superior performance of the gradient-enhanced version of the stochastic algorithm.

\subsubsection[Re = 800]{$Re = 800$}

\begin{figure}[H]
    \centering
    \subfloat[][Values of the minimum as function of the optimization iterations.]{\includegraphics[width=0.43\textwidth]{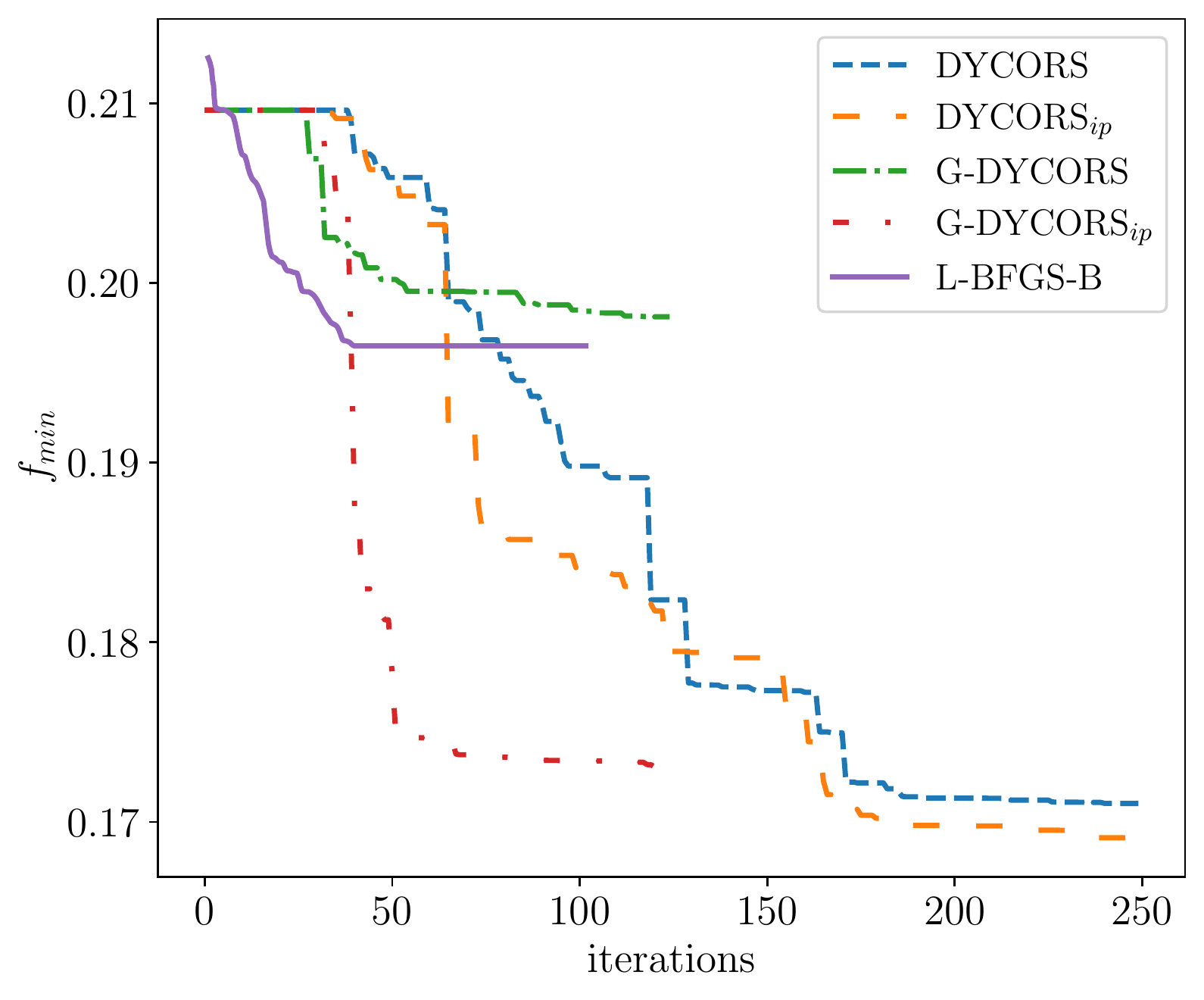} } \hfil
    \subfloat[][Values of the minimum as function of the computational cost.]{\includegraphics[width=0.43\textwidth]{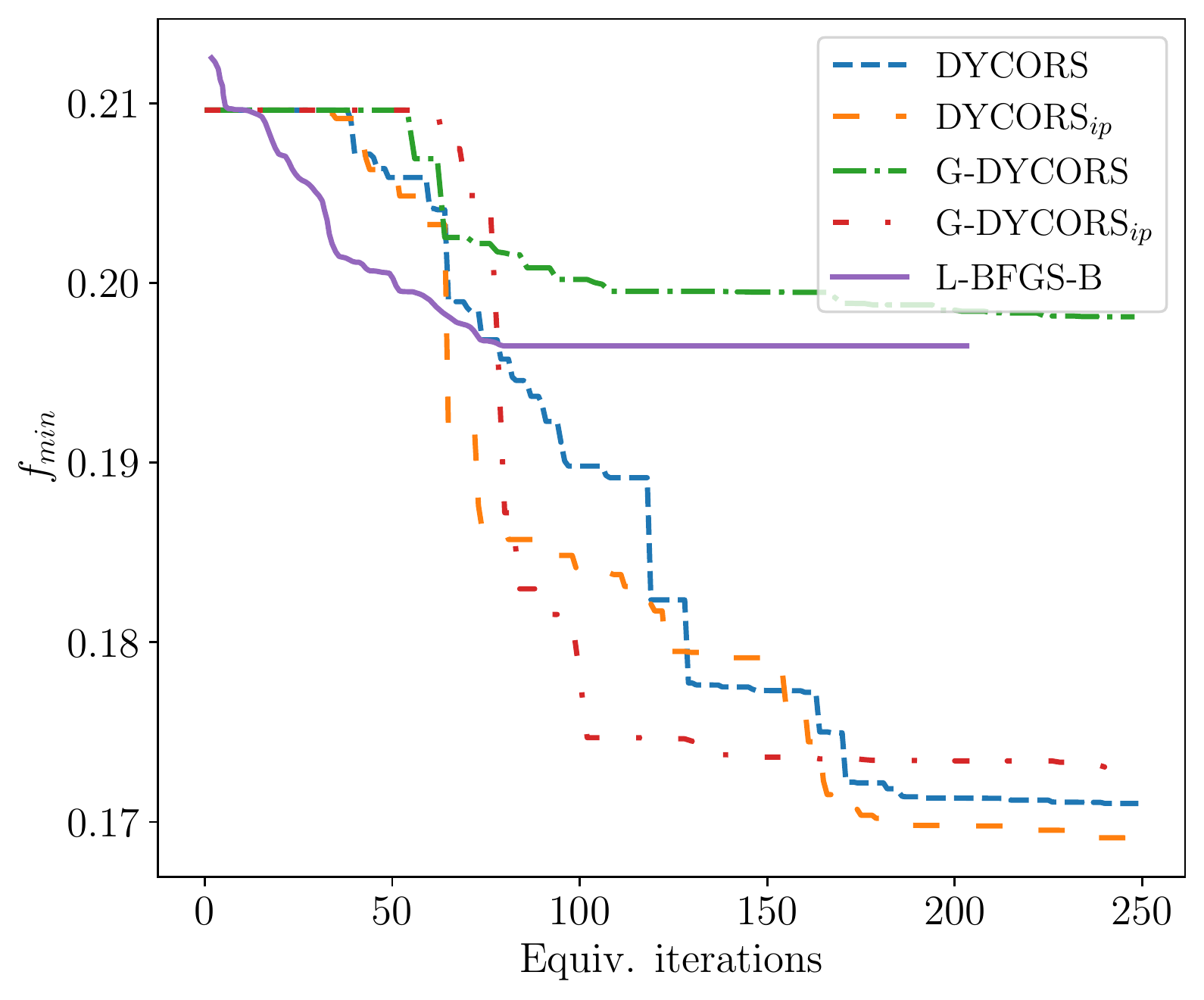} } \\
    \subfloat[][Optimal actuation obtained using the DYCORS and the G-DYCORS algorithms with and without optimizing the internal parameters of the kernel.]{\includegraphics[width=0.4\textwidth]{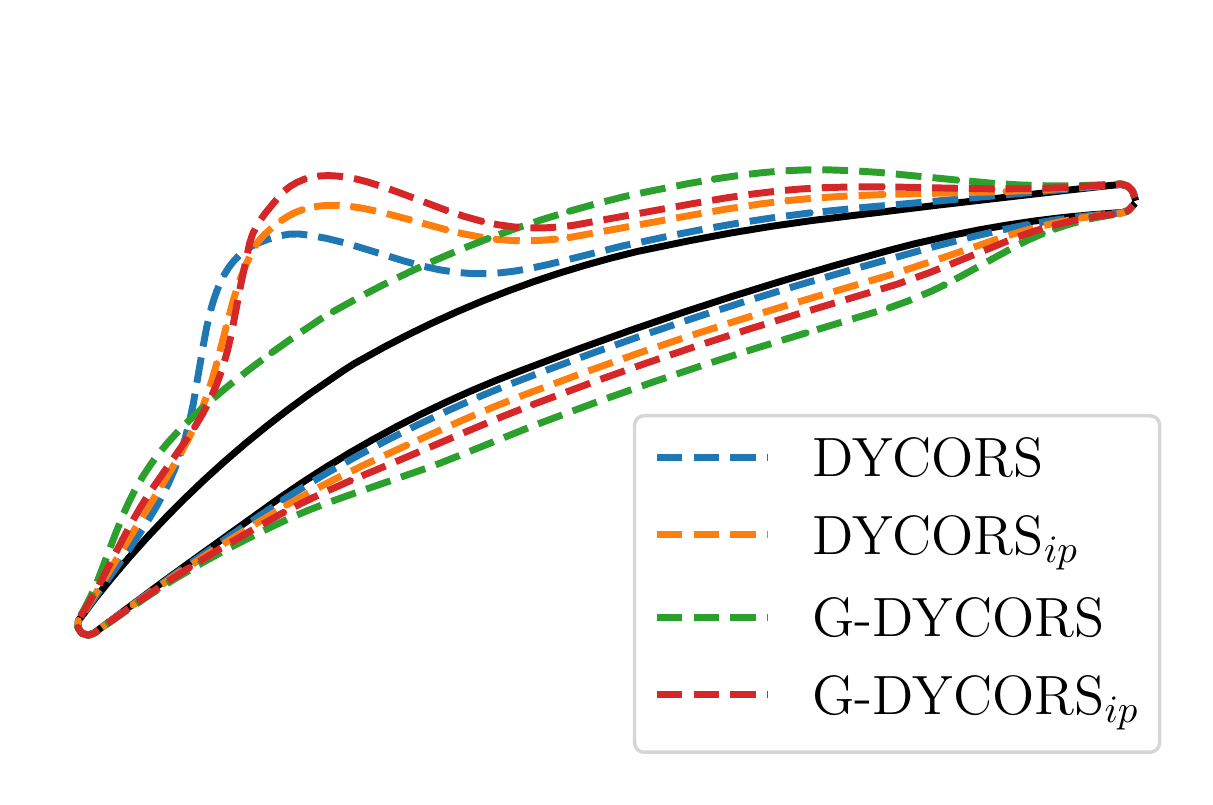} } \\
    \subfloat[][Average total pressure contour around the blades and total pressure profiles at the downstream measurement location for the case without actuation and the case using DYCORS$_{ip}$.]{\includegraphics[height=0.43\textwidth]{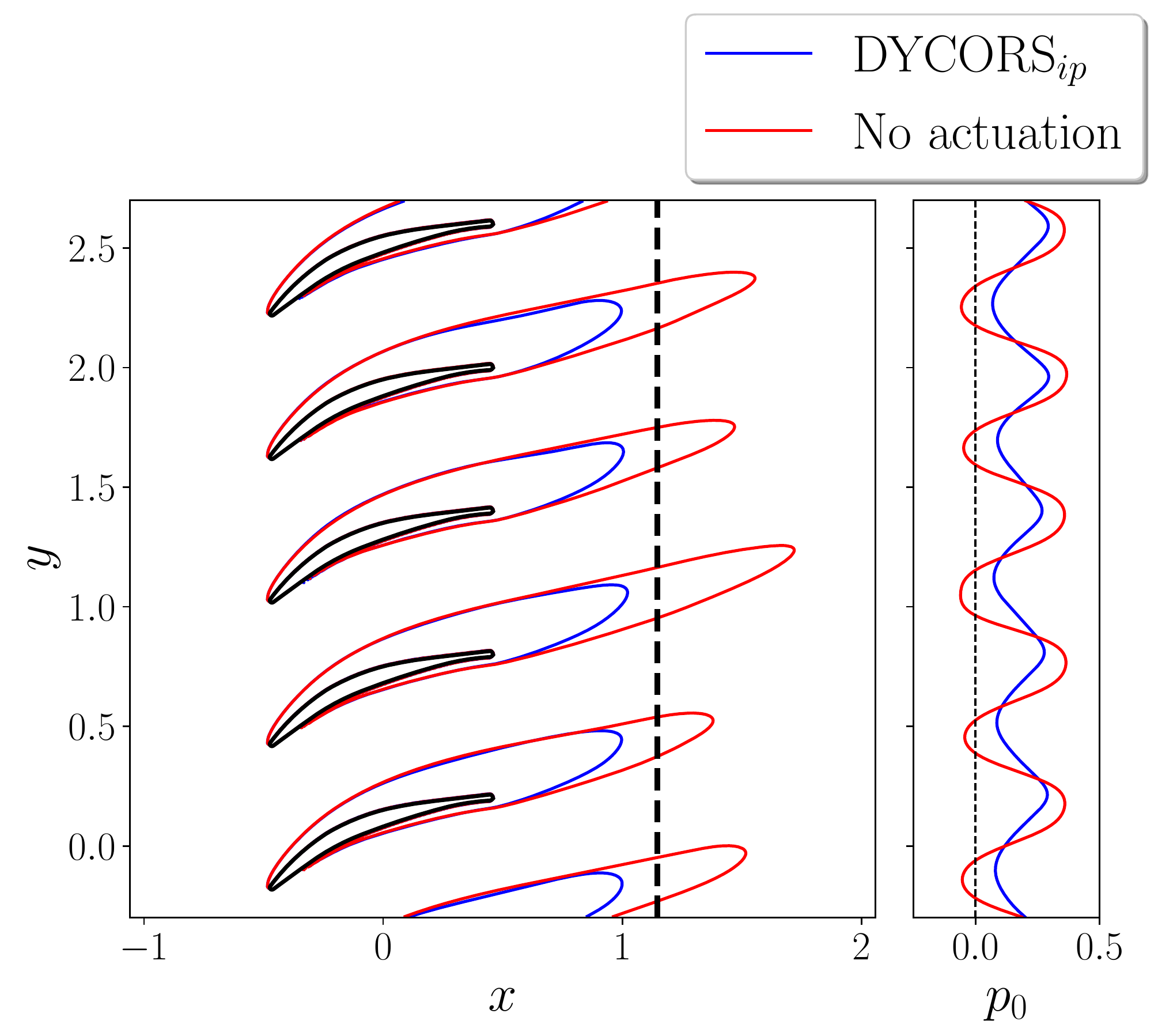} }
    \hfil
    \subfloat[][Instantaneous vorticity contour for the case without actuation and the case using DYCORS$_{ip}$.]{\includegraphics[height=0.35\textwidth]{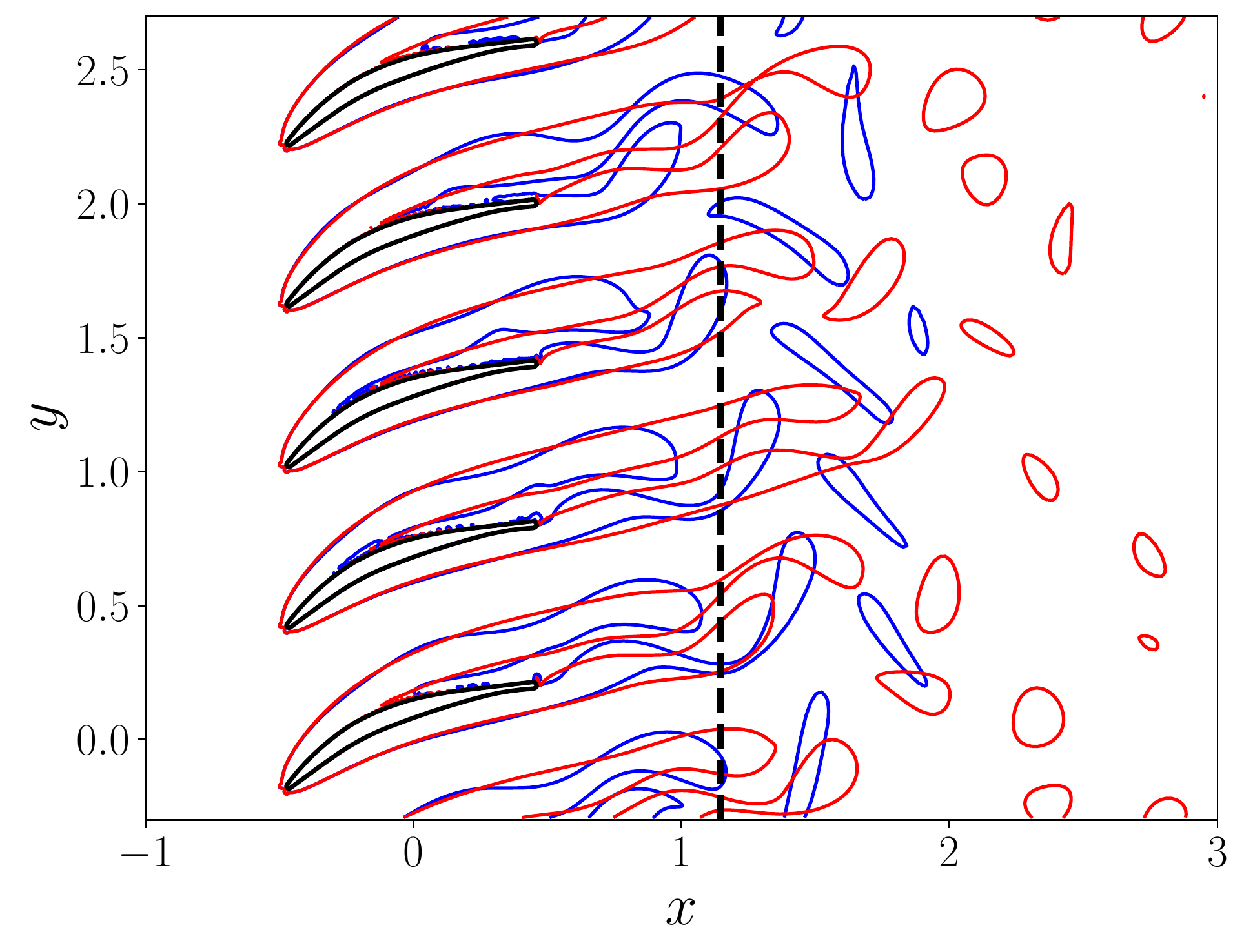} }
    \caption{Results at $Re = 800$.}
    \label{fig:comp_re800}
\end{figure}

Fig.~\ref{fig:comp_re800} displays the results obtained at $Re = 800$. First, Fig.~\ref{fig:comp_re800}(a) shows the convergence history of the objective function as a function of the number of iterations for each optimization performed at this Reynolds number. It can be seen that introducing the gradient in the surrogate model enhances the convergence rate of the algorithm as expected. This figure also demonstrates that an improvement is obtained when optimizing the internal parameters of the kernel, specially in the gradient enhanced version of the algorithm. Fig.\ref{fig:comp_re800}(b) shows the same convergence history plot but taking into account the computational cost of the optimization instead of the number of iterations performed. This is accomplished by multiplying the abscissa axis by a factor of $2$ in cases where the optimization is performed using the gradient information: G-DYCORS, G-DYCORS$_{ip}$, and L-BFGS-B algorithms. 

In order to find an explanation as to why the G-DYCORS$_{ip}$ algorithm did not achieve the best result at $Re=800$ we can examine the optimal actuators that were obtained with the different algorithms, shown in Fig.\ \ref{fig:comp_re800}(c). In this figure, the actuators at their maximum amplitude are plotted. It is clear that the DYCORS, DYCORS$_{ip}$ and G-DYCORS$_{ip}$ algorithms converged to a very similar solution in contrast to the G-DYCORS and the L-BFGS-B (not shown here) algorithms, which converged to a very different set of control parameters. This result suggest that the three algorithms arrived at a solution very close to the global minimum leaving little room for improvement. In addition, careful examination of the actuation profile shows that the profile is dominated by one actuator placed at the suction side of the blade between the leading edge and the mid-chord point, and the frequency of this actuator ($4.25$ rad/s in the case of the DYCORS$_{ip}$ and $4.43$ rad/s in the case of the G-DYCORS$_{ip}$) is roughly the same as the frequency of the instability ($4.76$ rad/s). This observation is also confirmed by looking at Fig.~\ref{fig:comp_re800}(d,e), which displays contours of the average total pressure field, the total pressure profile at the downstream measurement location and contours of the vorticity field at the last time step for the case with and without actuation (the actuation is ploted for the DYCORS$_{ip}$ algorithm). These figures show that a reduction in the size of the low total pressure region around the blades is obtained by decreasing the intensity of the vortical structures that are being generated. Moreover, in the optimized case all the averaged wakes present the same profile whereas this is not the case for the case without actuation. Optimal actuators corresponding to DYCORS and G-DYCORS$_{ip}$ present roughly the same flow fields as that of DYCORS$_{ip}$ (not shown here).

\subsubsection[Re = 2000]{$Re = 2000$}
\begin{figure}[H]
    \centering
    \subfloat[][Values of the minimum as function of the optimization iterations.]{\includegraphics[width=0.43\textwidth]{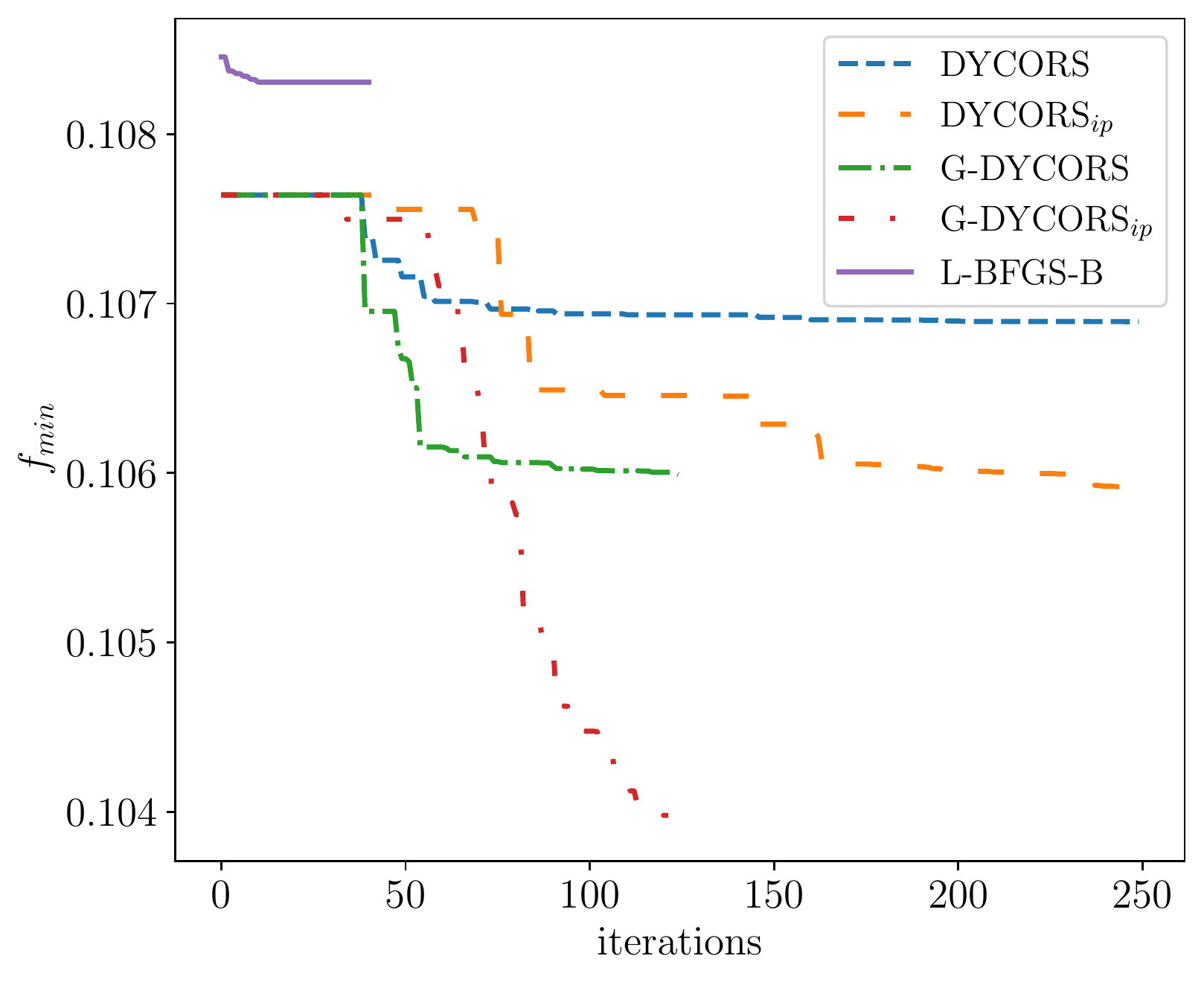} } \hfil
    \subfloat[][Values of the minimum as function of the computational cost.]{\includegraphics[width=0.43\textwidth]{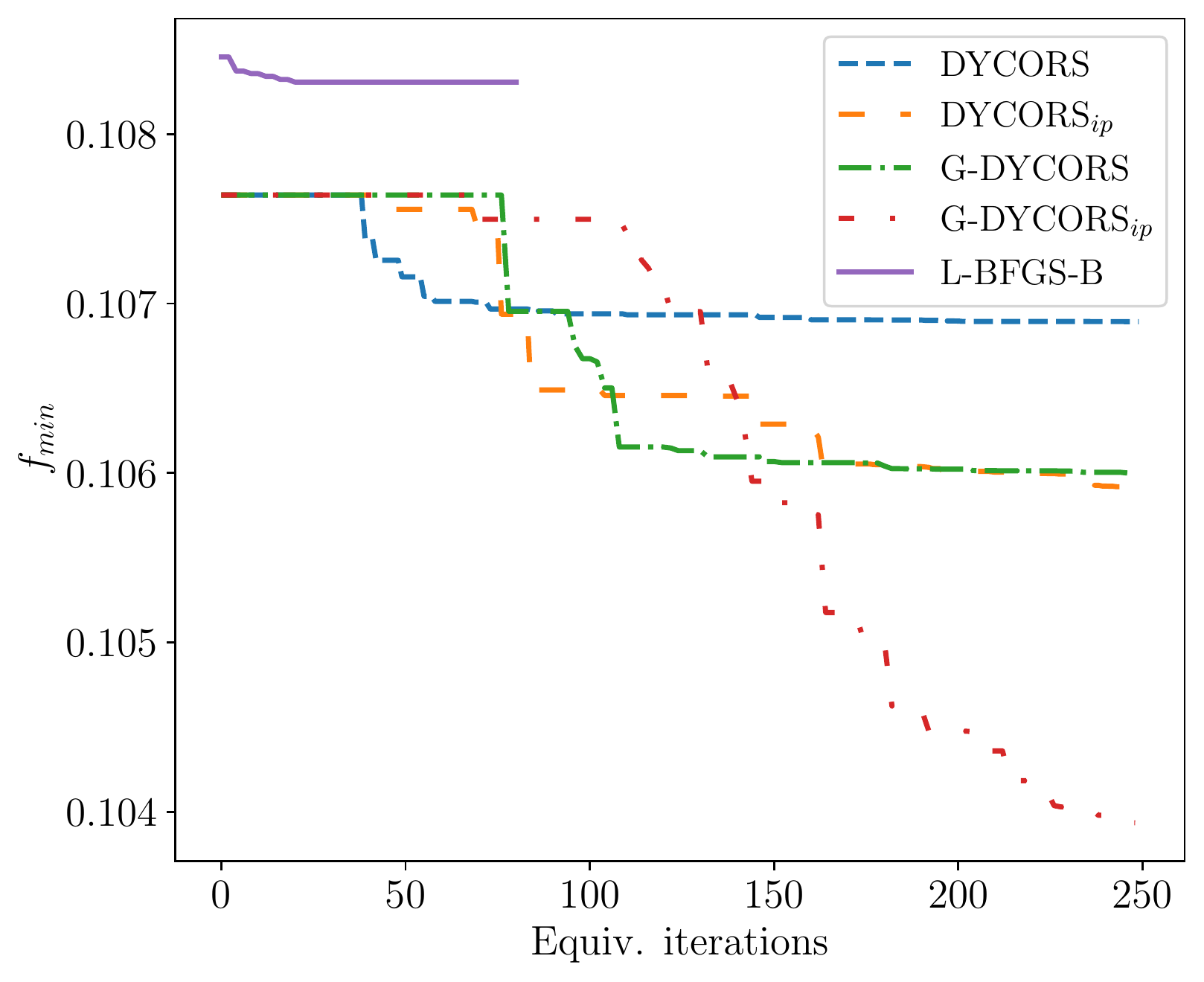} } \\
    \subfloat[][Optimal actuation obtained using the DYCORS and the G-DYCORS algorithms with and without optimizing the internal parameters of the kernel.]{\includegraphics[width=0.4\textwidth]{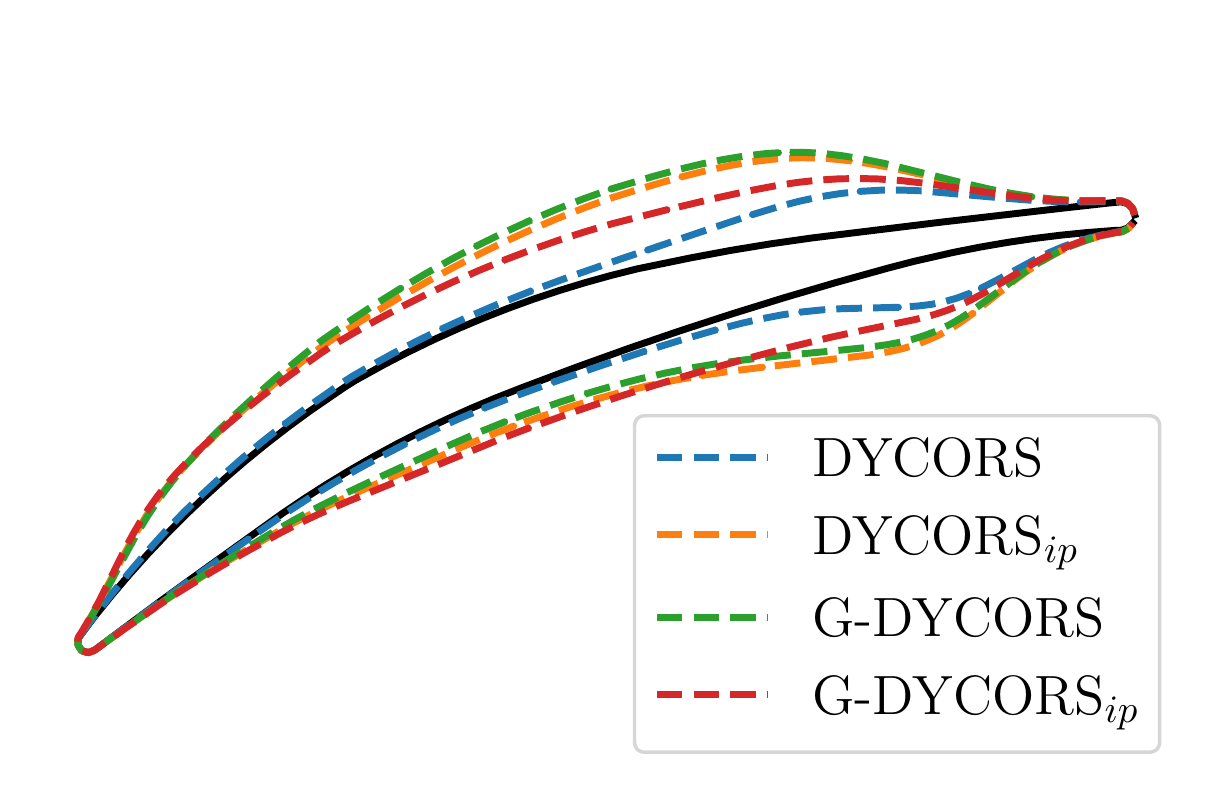} } \\
    \subfloat[][Average total pressure contour around the blades and total pressure profiles at the downstream measurement location for the case without actuation and the case using G-DYCORS$_{ip}$.]{\includegraphics[height=0.43\textwidth]{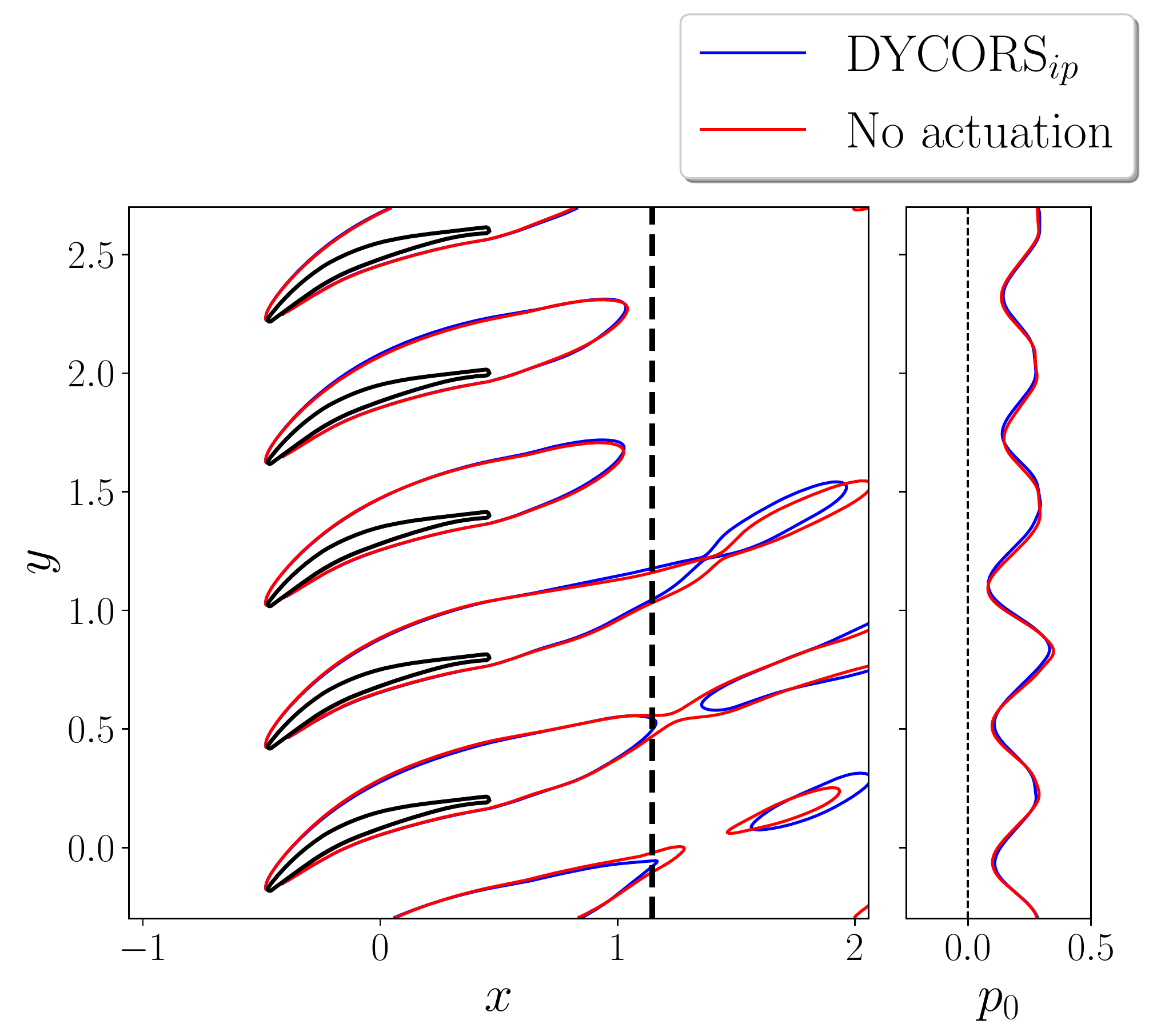} }
    \hfil
    \subfloat[][Instantaneous vorticity contour for the case without actuation and the case using G-DYCORS$_{ip}$.]{\includegraphics[height=0.35\textwidth]{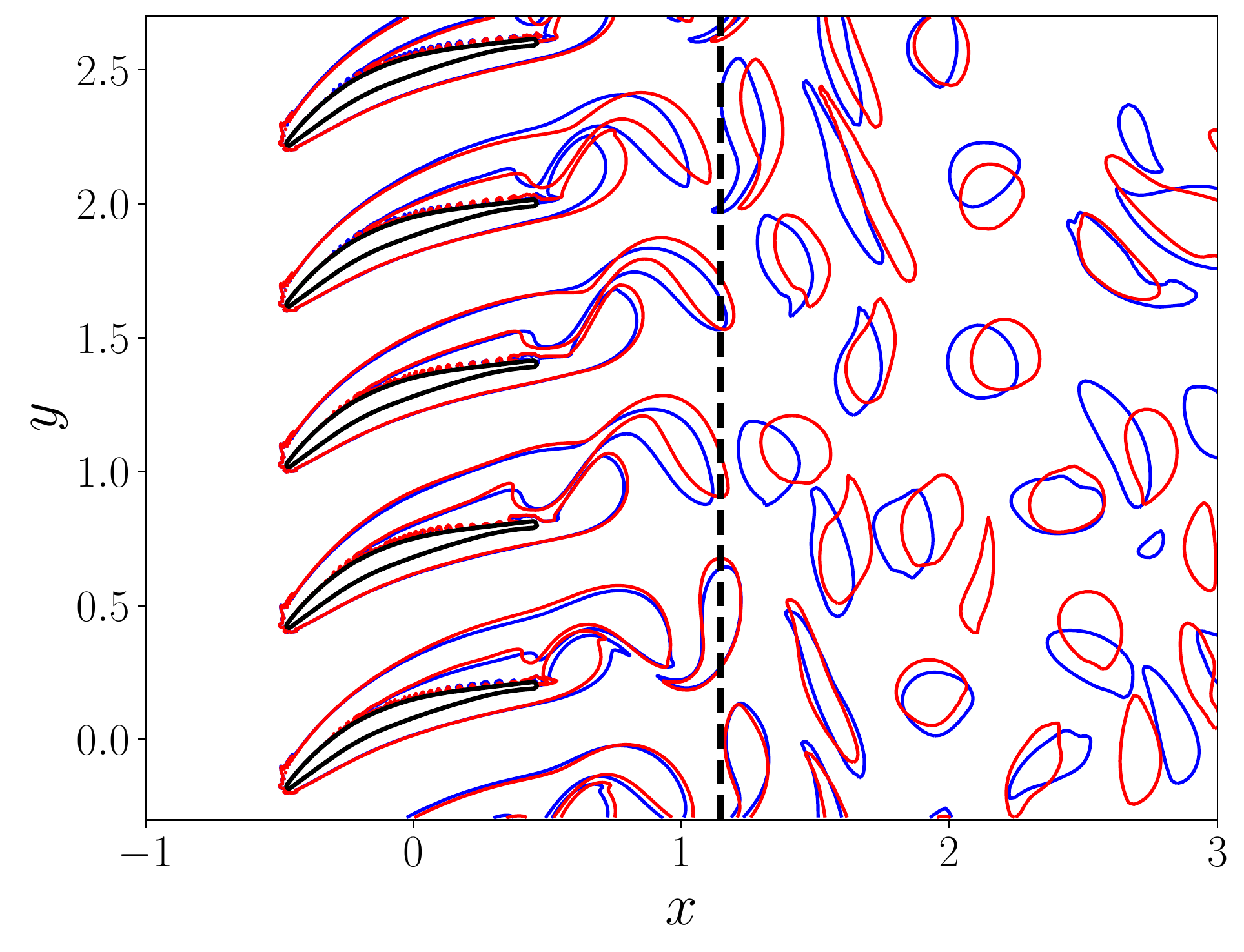} }
    \caption{Results at $Re = 2000$.}
    \label{fig:comp_re2000}
\end{figure}

The convergence history at $Re = 2000$ as function of the iterations and as function of the computational cost is shown in Fig.~\ref{fig:comp_re2000}(a,b). In this case, the G-DYCORS$_{ip}$ algorithm is the one that obtains the best result. Again, both versions where the internal parameters are optimized present better results than their counterparts without optimization. Also, both gradient-enhanced versions improve the convergence rate of the derivative-free versions. When taking into account the computational cost, the G-DYCORS$_{ip}$ algorithm is converged after 150 iterations.

The results of Table~\ref{tab:results}, suggest a considerable difference in the optimal actuators obtained with the G-DYCORS$_{ip}$ algorithms compared to the rest at $Re=2000$. However, Fig.~\ref{fig:comp_re2000}(c) demonstrates that this difference is small. In fact, comparing the reduction in the total pressure loss obtained with the optimal actuation to that of the case without actuation shows a smaller improvement at this Reynolds number than the rest. This result suggests that at this Reynolds number, the flow is not very sensitive to this type of actuation on the blade surface, and that this form of actuation is not the best strategy to reduce the total pressure loss. This can be deduced from the shape of the actuation profile as well, where no dominant actuator is selected, instead all the actuators have similar amplitudes.

The contours of the average total pressure, the total pressure profile at the measurement location and the contours of the vorticity field are plotted in Fig.~\ref{fig:comp_re2000}(d,e) for the case without actuation and for the optimal actuation obtained with the G-DYCORS$_{ip}$ algorithm at $Re=2000$. At this $Re$ number the instability is developing at the trailing edge of the blades instead of at the wake as in the case of $Re=800$. In this case there are no significant differences between the case without actuation and the optimized case as we already expected. Only in the wake of the 3 bottom blades the pressure contours show a small reduction in the size of the low total pressure region downstream of the cascade.

\subsubsection[Re = 4000]{$Re = 4000$}

\begin{figure}[H]
    \centering
    \subfloat[][Values of the minimum as function of the optimization iterations.]{\includegraphics[width=0.43\textwidth]{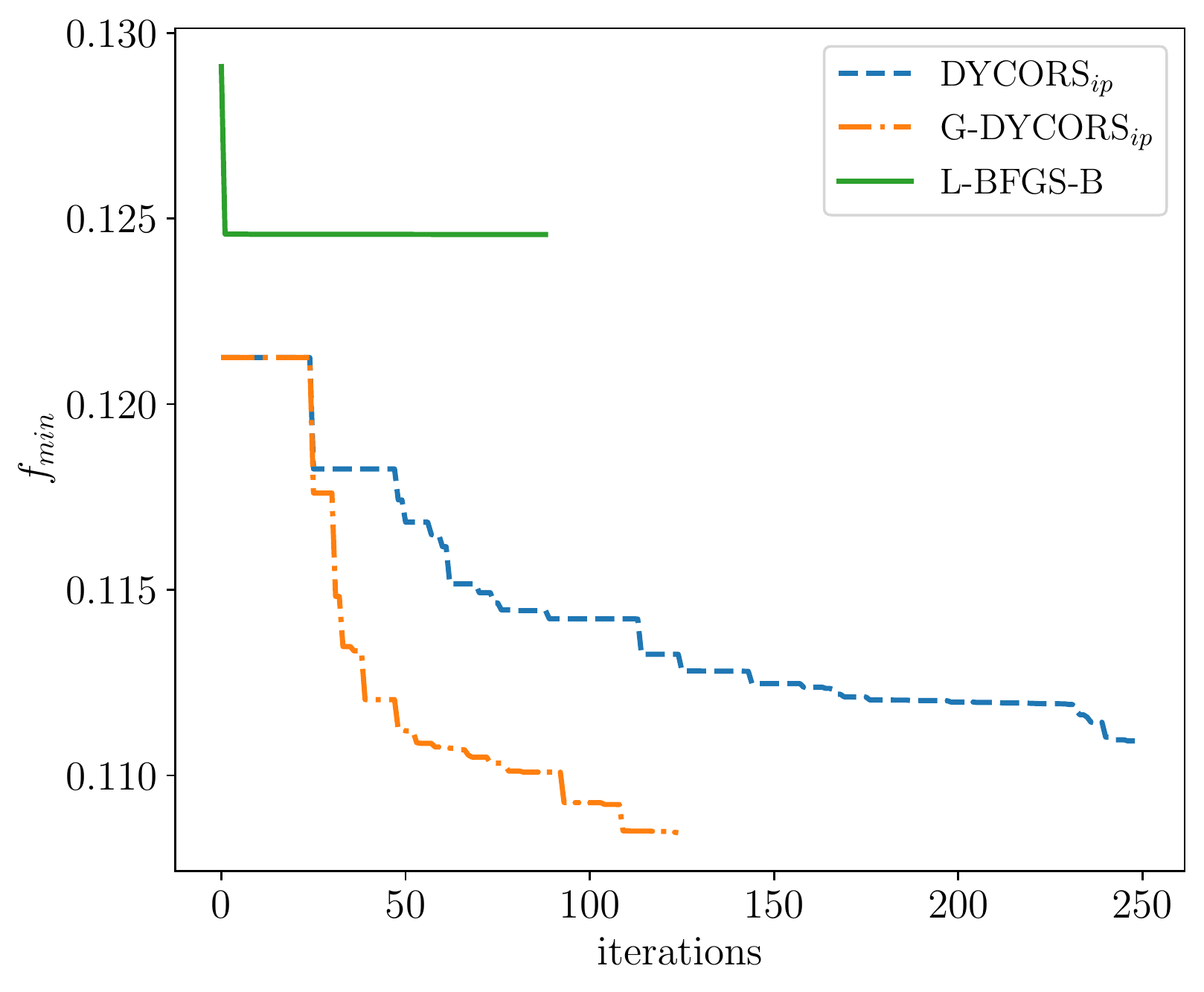} } \hfil
    \subfloat[][Values of the minimum as function of the computational cost.]{\includegraphics[width=0.43\textwidth]{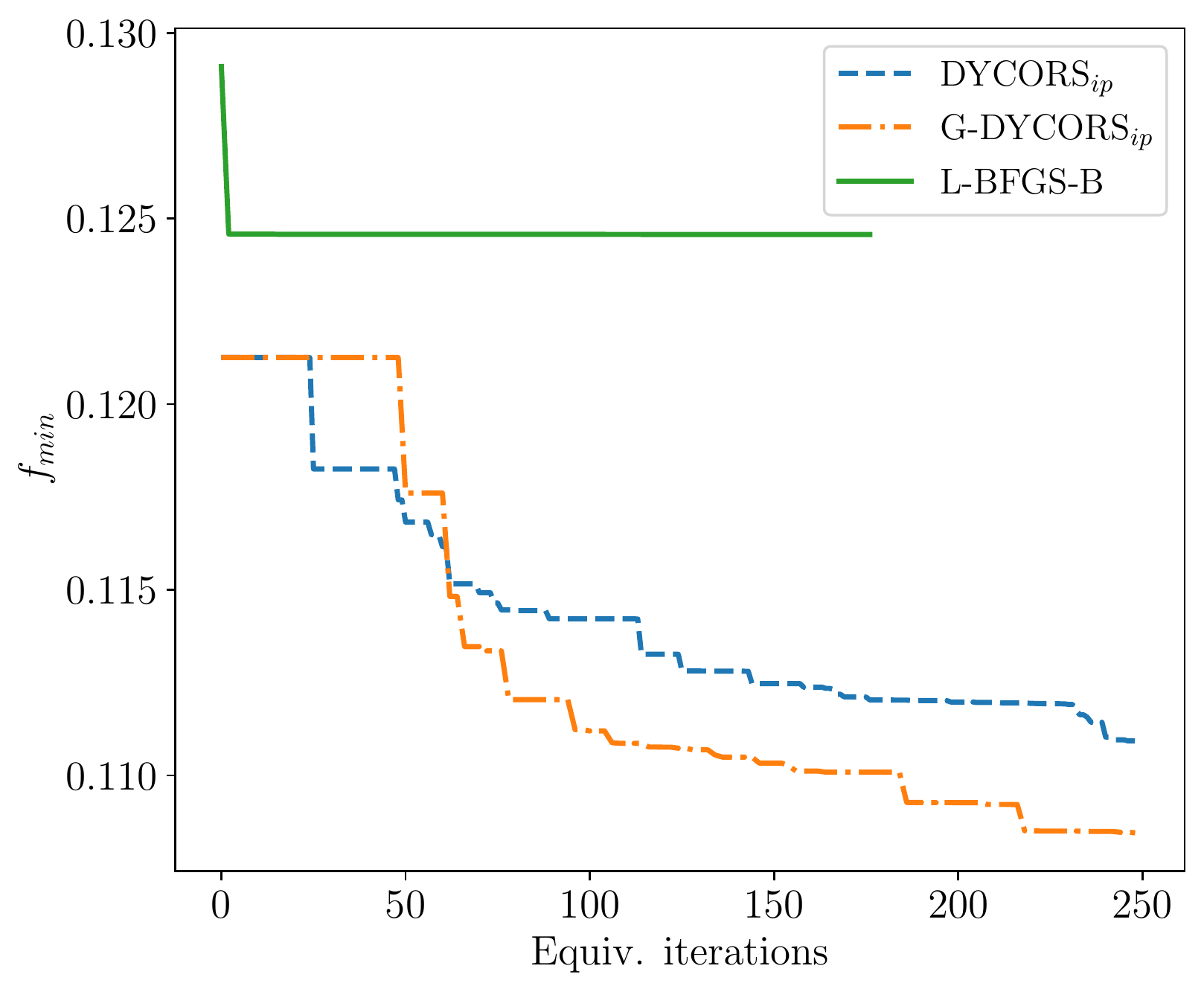} } \\
    \subfloat[][Optimal actuation obtained using the DYCORS and the G-DYCORS algorithms with and without optimizing the internal parameters of the kernel.]{\includegraphics[width=0.4\textwidth]{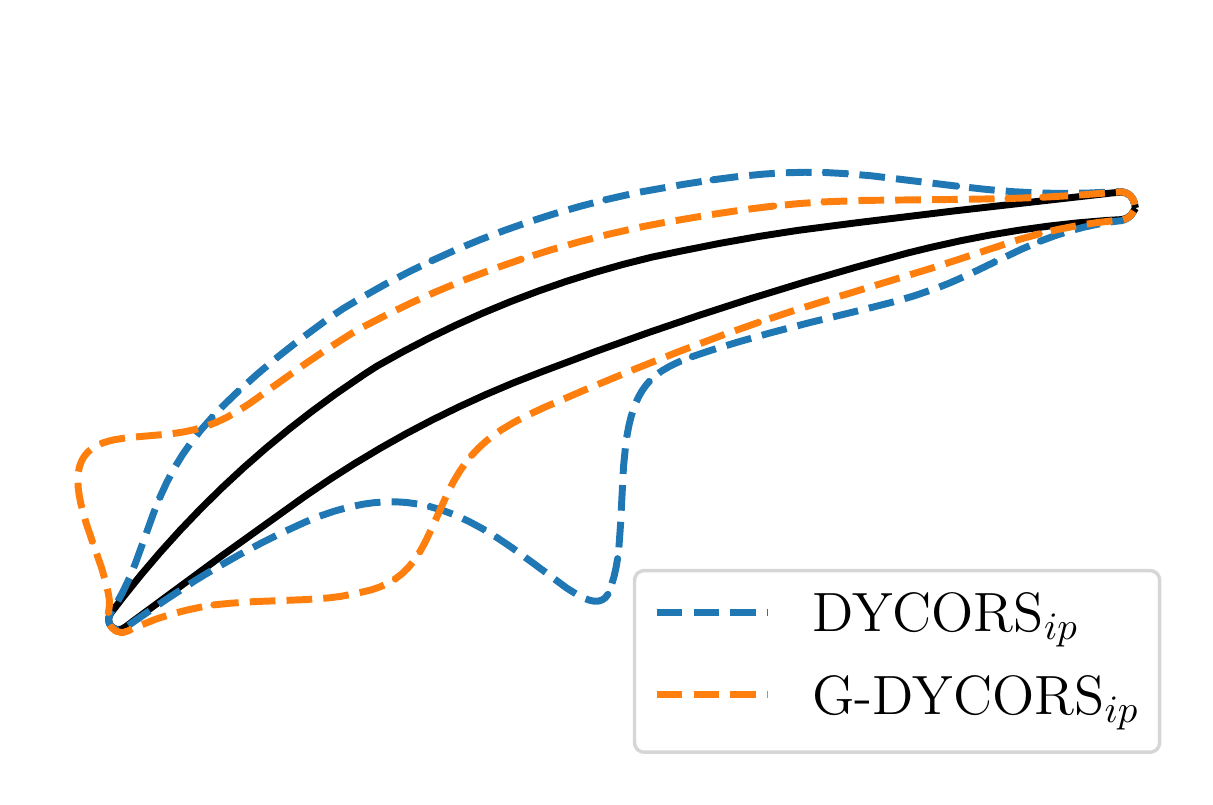} } \\
    \subfloat[][Average total pressure contour around the blades and total pressure profiles at the downstream measurement location for the case without actuation and the case using G-DYCORS$_{ip}$.]{\includegraphics[height=0.43\textwidth]{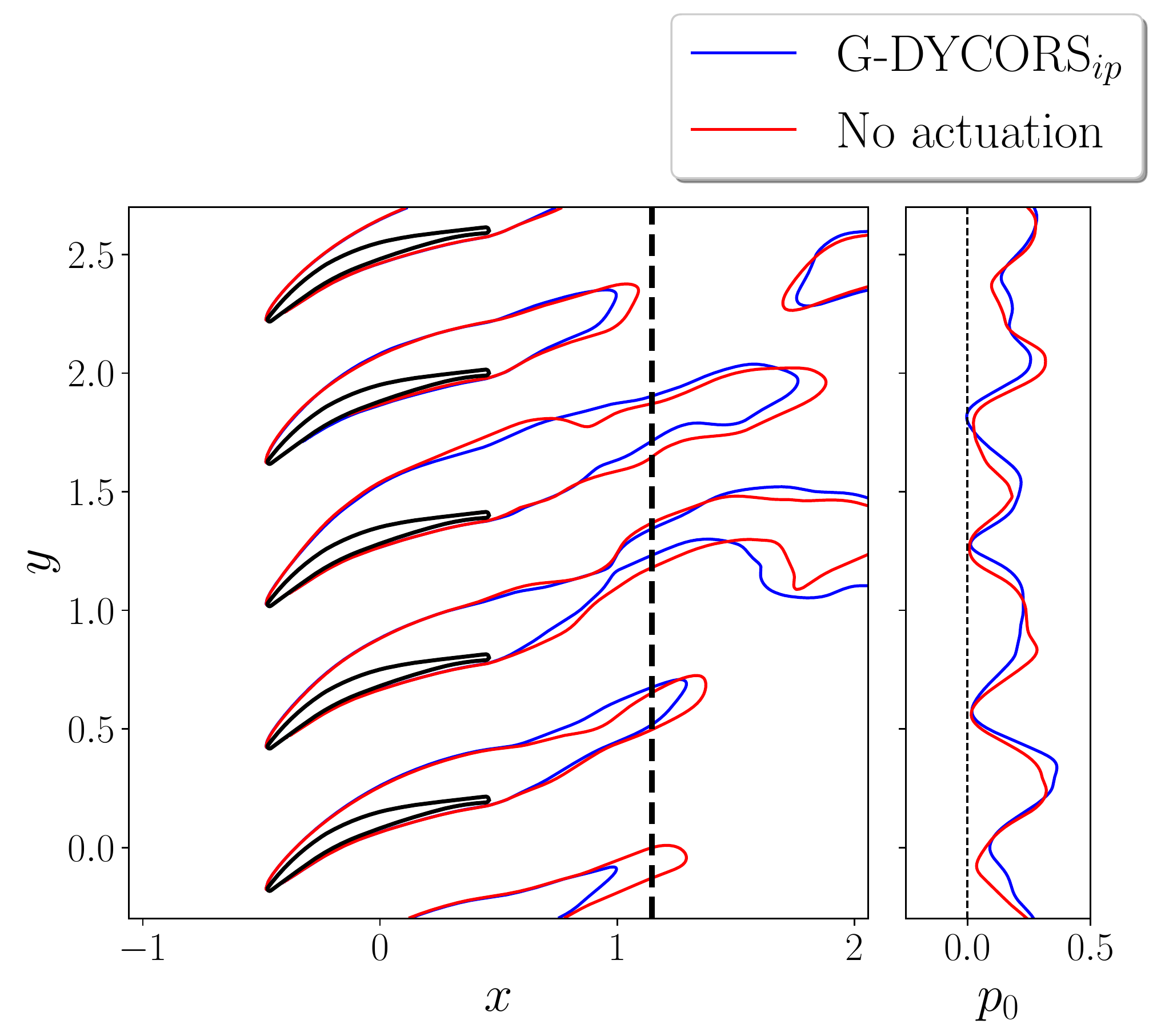} }
    \hfil
    \subfloat[][Instantaneous vorticity contour for the case without actuation and the case using G-DYCORS$_{ip}$.]{\includegraphics[height=0.35\textwidth]{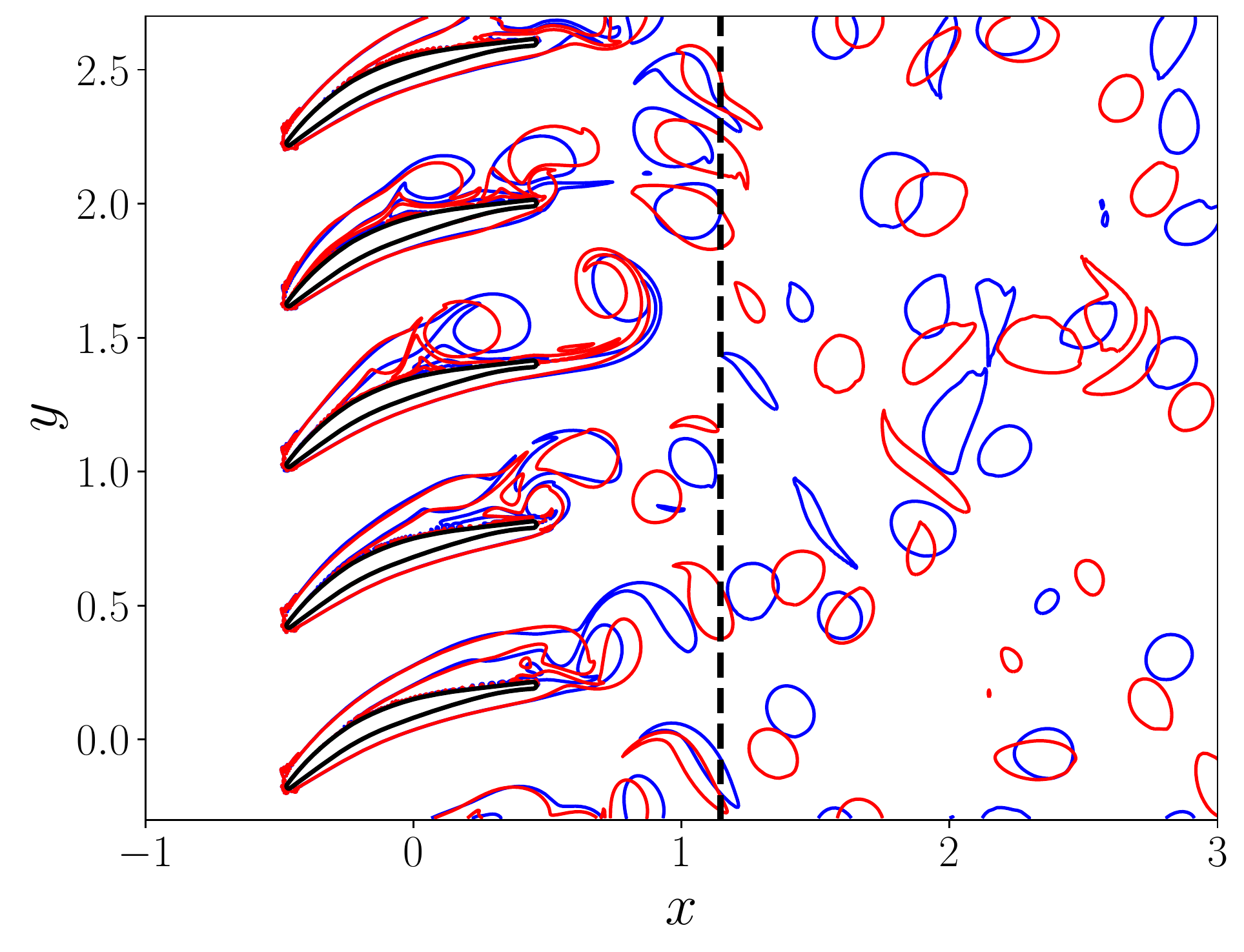} }
    \caption{Results at $Re = 4000$.}
    \label{fig:comp_re4000}
\end{figure}

At $Re=4000$, the optimal actuators obtained with the different algorithms present significant differences as shown by Fig.~\ref{fig:comp_re4000}(c), where the results of both stochastic optimizations are depicted. In the case of the DYCORS$_{ip}$ algorithm, the profile shows a dominant actuator in the pressure side close to the mid-chord point, while no dominant actuator exists in the suction side. Nevertheless, the overall contribution to the suction side suggests that the actuators placed on this side also have a greater influence on the minimization of the total pressure loss. In the case of the G-DYCORS$_{ip}$ algorithm, both the suction side and the pressure side are dominated by an actuator placed close to the leading edge, whereas the rest of the profile has smaller values of the tangential velocity when compared to the optimal actuator obtained by the derivative-free version of the algorithm. Although the DYCORS algorithm is ensured to achieve global convergence \cite{regis_stochastic_2007}, the large difference between the two results suggests that in this case they followed paths to different local minima. This can be explained also by Fig.~\ref{fig:comp_re4000}(a,b), where the convergence curves at this $Re$ number do not end in a plateau shape, suggesting that the algorithms did not reach the global minimum and more iterations of the optimization algorithm are required to reach this point. Nevertheless, the convergence curves show that the G-DYCORS$_{ip}$ is able to reach the same value of the objective function employing half the computational cost of the DYCORS$_{ip}$ algorithm.

Considering Fig.~\ref{fig:comp_re4000}(d,e), where the contours of the average total pressure field, the total pressure profile at the measurement location and the contours of the vorticity field are illustrated for the case without actuation and for the optimal actuation obtained with the G-DYCORS$_{ip}$, it can be observed that the vortical structures are being generated on the suction side of the blades. This explains why the dominant actuators are place so close to the leading edge in the optimal case. This way the actuation is capable of disturbing the shedding of these vortices and hence modifying the total pressure downstream of the blades. It can be seen that the shed vortices are not noticeably different qualitatively between the two cases, however the vortical structures in the wake show more variations, suggesting that the actuation is not changing the intensity of the vortices but their interaction. Again, the total pressure contours show larger low total pressure regions at the measurement location for the case without actuations although the differences are not as large as in the case at $Re = 800$.

\section{Summary and conclusions}
\label{sec:conclusions}
In this work we have developed an enhanced version of the derivative-free stochastic DYCORS algorithm by performing the optimization of the internal parameters of the kernel. An alternative version of the algorithm is also proposed by adding gradient information into the surrogate model to create a gradient-enhanced version of the original algorithm, which may be useful whenever the gradient information can be obtained. These two modifications improve the accuracy of the surrogate model and therefore improve the convergence rate of the algorithm. To optimize the internal parameters, the leave-one-out error is used in the case of the derivative-free version. In the case of the gradient-enhanced version, it has been found that the leave-one-out error presented in \cite{bompard_surrogate_2010} does not perform satisfactorily and therefore an alternative method is proposed to optimize the values of the internal parameters based on the values of the gradients at the evaluated points. An implementation of both DYCORS and G-DYCORS algorithms together with brief documentation of the code is available at \cite{quiros_rodriguez_dycors_2022}.

We have analyzed the performance of the stochastic algorithms and have compared it to the performance of the commonly used gradient-based algorithm L-BFGS-B at different flow regimes, from flows at roughly the critical Reynolds number to flows exhibiting chaotic behaviour. In all the cases, optimizing the internal parameters of the kernel has significantly improved the convergence rate of the algorithm. Moreover, the gradient-enhanced version has clearly outperformed the derivative-free version in two out of the three cases that have been analyzed. The comparison with the gradient-based algorithm L-BFGS-B has demonstrated that stochastic algorithms are able to achieve better results even at low $Re$ numbers where the flows exhibits a purely periodic behaviour and where the objective function is not expected to present many local minima.

The convergence plots show that the gradient-enhanced version of the algorithm always presents a better convergence rate than the derivative-free version, even when the cost of evaluating the gradient has been factored in. However, at the lowest $Re$ number studied, taking into account the computational cost of computing the gradient, both versions of the algorithm perform similarly. In order to compute the gradient information we have made use of the adjoint method, therefore the cost of performing a gradient evaluation is roughly the same as the cost of evaluating the objective function. Nevertheless, there exist methods that can improve the cost of gradient extraction e.g. the parallel-in-time method \cite{skene_parallel--time_2020, costanzo_parallel--time_2022} where the linear equations are partitioned by separating the homogeneous and inhomogeneous parts of the equations, resulting in a speeds-up of the computation. Reducing the computational cost of performing a gradient evaluation would improve the performance of the gradient-enhanced version of the algorithm compared to the derivative-free version.

Finally, the gradient-enhanced version can still be improved by updating the function employed to generate the trial points. Since in this case the gradient is evaluated at the best evaluated point, a skew-normal distribution can be used instead of a symmetric normal distribution to randomly generate the trial points by taking into account the direction where the gradient is pointing. This would accelerate convergence to local minima and therefore improve the overall performance of the algorithm and will be investigated further in the future. 

\section*{Acknowledgements}

We gratefully acknowledge the computer resources at Finisterrae and the technical support provided by CESGA (Grant No. IM-2020-3-0020 and IM-2021-1-0021).

\appendix
\section{Adjoint-based gradient computation}
\label{sec:app_adjoint}

By linearizing the system of equations \eqref{eq:discretized-eqs} about a steady baseflow, we obtain the following linear system of equations for the advancement of small perturbations $\bm{\mathsf{q}}$
\begin{equation} \label{eq:linear-eqs}
    \begin{pmatrix}
    \bm{\mathsf{A}} & \bm{\mathsf{Q}} \\ 
    \bm{\mathsf{Q}}^T & \bm{\mathsf{0}}
    \end{pmatrix} \begin{pmatrix}
    \bm{\mathsf{q}}^{k+1} \\ 
    \boldsymbol{\lambda}
    \end{pmatrix} = \begin{pmatrix}
    (\bm{\mathsf{B}} -\frac{3}{2} \bm{\mathsf{N}})\bm{\mathsf{q}}^k + \frac{1}{2} \bm{\mathsf{N}}(\bm{\mathsf{q}}^{k-1}) \\ 
    \mathbf{0}
    \end{pmatrix},
\end{equation}
where the matrix $\bm{\mathsf{N}}$ represents the linearized advection operator. This linearized system of equations is used to perform the gradients computation.

In order to obtain the desired derivatives of the objective function $\mathcal{J}$, we make use of adjoint variables to efficiently compute the gradients by solving a system of equations with a similar computational cost than the cost of the forward simulation. By introducing the governing equations $R$ as a constraint in the cost function and using Lagrange multipliers we can transform the minimization problem
\begin{equation}
    \min \mathcal{J}(\mathbf{q}^0,\dots,\mathbf{q}^K,\bx)\ \mathrm{s.t.}\  R(\mathbf{q}^{i-2}, \mathbf{q}^{i-1}, \mathbf{q}^{i}, \bx) = 0\ \forall\ i \in \{ 1,\dots,K \},
\end{equation}
into an unconstrained problem
\begin{equation}
    \min \mathcal{L}(\mathbf{q}^0,\dots,\mathbf{q}^K,\bx,\lambda) = \mathcal{J}(\mathbf{q}^0,\dots,\mathbf{q}^K,\bx) - \sum_{i=1}^{K} \lambda_i^T R(\mathbf{q}^{i-2}, \mathbf{q}^{i-1}, \mathbf{q}^{i}, \bx),
\end{equation}
where $\lambda_i$ is the Lagrange multiplier corresponding to the residual of the $i$-th time step and $\mathcal{L}$ is the new cost function. We are interested in computing the gradients of the objective function with respect to the control parameters
\begin{equation}
    \dfrac{\mathrm{d}\mathcal{J}}{\mathrm{d}\bx} = \pder{\mathcal{J}}{\bx} + \sum_{i=0}^{i=K} \pder{\mathcal{J}}{\mathbf{q}^{i}} \pder{\mathbf{q}^i}{\bx},
\end{equation}
which by employing the first-order optimality conditions can be rewritten as 
\begin{equation}
    \dfrac{\mathrm{d}\mathcal{J}}{\mathrm{d}\bx} = \pder{\mathcal{J}}{\bx} - \sum_{i=0}^{i=K}\lambda_i^T \pder{R}{\bx},
\end{equation}
where the Lagrange multipliers $\lambda_i$ are obtained by solving the adjoint system backwards in time given by
\begin{equation}
\begin{split}
    &\pder{\mathcal{J}}{\mathbf{q}^0} = 2\lambda_1^T \pder{R(\mathbf{q}^0, \mathbf{q}^{0}, \mathbf{q}^{1}, \bx)}{\mathbf{q}^0} + \lambda_2^T \pder{R(\mathbf{q}^0, \mathbf{q}^{1}, \mathbf{q}^{2}, \bx)}{\mathbf{q}^0} \\
    &\pder{\mathcal{J}}{\mathbf{q}^1} = \lambda_1^T \pder{R(\mathbf{q}^0, \mathbf{q}^{0}, \mathbf{q}^{1}, \bx)}{\mathbf{q}^1} + \lambda_2^T \pder{R(\mathbf{q}^0, \mathbf{q}^{1}, \mathbf{q}^{2}, \bx)}{\mathbf{q}^1} + \lambda_3^T \pder{R(\mathbf{q}^1, \mathbf{q}^{2}, \mathbf{q}^{3}, \bx)}{\mathbf{q}^1} \\
    &\pder{\mathcal{J}}{\mathbf{q}^i} = \lambda_i^T \pder{R(\mathbf{q}^{i-2}, \mathbf{q}^{i-1}, \mathbf{q}^{i}, \bx)}{\mathbf{q}^i} + \lambda_{i+1}^T \pder{R(\mathbf{q}^{i-1}, \mathbf{q}^{i}, \mathbf{q}^{i+1}, \bx)}{\mathbf{q}^i} + \lambda_{i+2}^T \pder{R(\mathbf{q}^i, \mathbf{q}^{i+1}, \mathbf{q}^{i+2}, \bx)}{\mathbf{q}^i}\ \forall\ i \in \{ 2,\dots,K-2 \} \\
    &\pder{\mathcal{J}}{\mathbf{q}^{K-1}} = \lambda_{K-1}^T \pder{R(\mathbf{q}^{K-3}, \mathbf{q}^{K-2}, \mathbf{q}^{K-1}, \bx)}{\mathbf{q}^{K-1}} + \lambda_{K}^T \pder{R(\mathbf{q}^{K-2}, \mathbf{q}^{K-1}, \mathbf{q}^{K}, \bx)}{\mathbf{q}^{K-1}} \\
    &\pder{\mathcal{J}}{\mathbf{q}^{K}} = \lambda_{K}^T \pder{R(\mathbf{q}^{K-2}, \mathbf{q}^{K-1}, \mathbf{q}^{K}, \bx)}{\mathbf{q}^{K}}.
\end{split}
\end{equation}

\section{Actuation details}
\label{sec:app_actuation}

The function $f$ that appears in Eq.\ \eqref{eq:actuation} is a Gaussian-like function periodic over $2 \pi$ given by
\begin{equation}
    \begin{split}
        f(\theta, \theta_i, \sigma_i) &= f_1(\theta, 0) f_1(\theta, \pi) f_1(\theta, 2\pi) \sum_{n=-\infty}^\infty \frac{1}{\sqrt{2\pi}\sigma_i} e^{\displaystyle-\frac{1}{2}\left(\frac{\theta-\theta_i - 2\pi n}{\sigma_i}\right)^2} \\ 
        &= f_1(\theta, 0) f_1(\theta, \pi) f_1(\theta, 2\pi) \frac{1}{2\pi} \vartheta_3 \left( \frac{1}{2}\left(\theta-\theta_i\right), e^{\displaystyle-\frac{\sigma_i^2}{2}}\right),
    \end{split}
\end{equation}
where $\vartheta_3$ is the Jacobi theta function. By considering this function we ensure that a continuous distribution of tangential velocity will be obtained when applied to the blade surface. The function $f_1$ damps the tangential velocity to 0 close to the leading and trailing edges in order to avoid problems related to the high curvature of these sections of the blade. It is given by,
\begin{equation}
    f_1 (\theta, \alpha) = \left( 1-\exp{\left( -\dfrac{(\theta-\alpha)^2}{\sigma_d} \right)} \right),
\end{equation}
where $\sigma_d = 0.1$ is chosen so that the damped region does not extend much throughout the blades surface.

\section{Algorithms}
\label{sec:app_algo}

This appendix presents the algorithms that have been previously mentioned in the paper. The Algorithm \ref{algo:dycors} describes the steps that need to be performed in order to generate and evaluate the new trial points. The Algorithm \ref{algo:snp} provides the steps to select the next point to be evaluated using the expensive function evaluation given the set of trial points already evaluated at the surrogate surface. Finally, the Algorithm \ref{algo:update} presents the steps that need to be carried out to update several parameters needed in the optimization procedure.

\begin{algorithm}
    \DontPrintSemicolon
    \SetAlgoLined
    \KwIn{
        Current iteration, $n$ \newline
        Number of trial points, $k$ \newline
        Standard deviation, $\sigma_n$\newline
        Interpolant, $s_n$ \newline
        Evaluated points, $\mathcal{A}_n$\newline
        Coefficients of the surrogate model, $\boldsymbol{\lambda}$ and $\mathbf{c}$ \newline
    }
    \KwResult{Set of values of the trial points evaluated at the surrogate model, $\mathcal{B}_n$} \;
    
    \textit{\textbf{Compute probability of perturbing a coordinate}}: $p_\mathrm{pert}=\varphi(n)$ using Eq.\ \eqref{eq:phi}\;
    
    \textit{\textbf{Select coordinates to perturb}}: $\mathcal{I}_\mathrm{pert} = \left\lbrace i:w_i<p_\mathrm{pert} \right\rbrace $, where $w_i$ for $ i=1,\dots,d$ are generated randomly
        \SetAlgoNoLine\DummyBlock{\SetAlgoLined\If{$\mathcal{I}_\mathrm{pert}=\emptyset$}{
            $\mathcal{I}_\mathrm{pert}=\left\lbrace j \right\rbrace$ where $j$ is selected randomly from $\left\lbrace 1,\dots,d \right\rbrace$}}
    
    \textit{\textbf{Generate trial points}}: $\textbf{y}_{n,j} = \bx_\mathrm{best}+\textbf{z}_j$ for $ j=1,\dots,k$, where $\textbf{z}_j^i=0\ \forall\ i\not\in \mathcal{I}_\mathrm{pert}$ and a random number from the normal distribution $\mathcal{N}(0,\sigma_n^2)$\;
    
    \textit{\textbf{Ensure trial points are in the domain}}:
    \SetAlgoNoLine\DummyBlock{\SetAlgoLined\If{$\textbf{y}_{n,j} \not\in \mathcal{D}$}{
        Replace $\textbf{y}_{n,j}$ by closest point in the boundary $\partial\mathcal{D}$}}
    
    \textit{\textbf{Evaluate trial points}}: Compute $\mathcal{B}_n = s_n(\by_{n,j}, \mathcal{A}_n, \boldsymbol{\lambda}, \textbf{c})$ for $j=1,\dots,k$ where $s_n$ is given by Eqs.\ \eqref{eq:interp_RBF} and \eqref{eq:interp_GRBF} \;
    
    \caption{\textbf{Function} trial\_points()}
    \label{algo:dycors}
\end{algorithm}

\begin{algorithm}
    \DontPrintSemicolon
    \SetAlgoLined
    \KwIn{Current iteration, $n$\newline
        Weight pattern, $\boldsymbol{\Upsilon}$\newline
        Evaluated points, $\mathcal{A}_n$\newline
        Set of trial points, $\textbf{y}_n$\newline
        Set of values of the trial points, $\mathcal{B}_n$
    }
    \KwResult{Best candidate point, $\bx_{n+1}$} \;
    
    \textit{\textbf{Compute RBF score}}: Compute $s_n^\mathrm{max} = \max (\mathcal{B}_n)$ and $s_n^\mathrm{min} = \min (\mathcal{B}_n)$ 
    \SetAlgoNoLine\DummyBlock{\SetAlgoLined\eIf{$s_n^\mathrm{max} = s_n^\mathrm{min}$}{$V_{n,j}^\mathrm{RBF} = 1$ for $j=1,\dots,k$}{$V_{n,j}^\mathrm{RBF} = (\mathcal{B}_{n,j}-s_n^\mathrm{min})/(s_n^\mathrm{max} - s_n^\mathrm{min})$ for $j=1,\dots,k$}}
    
    \textit{\textbf{Compute distance score}}: Compute $\Delta_{n,j} = \min \left \| \textbf{y}_{n,j} - \bx \right \| : \bx\in\mathcal{A}_n $, $\Delta_n^\mathrm{max} = \max (\Delta_{n,j})$ and $\Delta_n^\mathrm{min} = \min (\Delta_{n,j})$ for $j=1,\dots,k$ 
    \SetAlgoNoLine\DummyBlock{\SetAlgoLined\eIf{$\Delta_n^\mathrm{max} = \Delta_n^\mathrm{min}$}{$V_{n,j}^\mathrm{dist} = 1$ for $j=1,\dots,k$}{$V_{n,j}^\mathrm{dist} = (\Delta_n^\mathrm{max}-\Delta_{n,j})/(\Delta_n^\mathrm{max} - \Delta_n^\mathrm{min})$ for $j=1,\dots,k$}}
    
    \textit{\textbf{Pick weights}}: $w_n^\mathrm{RBF} = \Upsilon_{n\%4}$ and $w_n^\mathrm{dist} = 1-w_n^\mathrm{RBF}$\;
    
    \textit{\textbf{Compute global score}}: $w_{n,j} = w_n^\mathrm{RBF} V_{n,j}^\mathrm{RBF} + w_n^\mathrm{dist}V_{n,j}^\mathrm{dist} $\;
    
    \textit{\textbf{Select next point to be evaluated}}: $ \bx_{n+1} = \min_{\textbf{y}_{n,j}} w_{n,j}$\;
    
    \caption{\textbf{Function} select\_next\_point()}
    \label{algo:snp}
\end{algorithm}

\begin{algorithm}
    \DontPrintSemicolon
    \SetAlgoLined
    
    \KwIn{Current iteration, $n$\newline
        Current minimum, $\bx_\mathrm{best}$ and $f_\mathrm{best}$\newline
        Last evaluated point, $\bx_{n+1}$ and $f(\bx_{n+1})$\newline
        Counters, $C_\mathrm{f}$ and $C_\mathrm{s}$\newline
        Counter limits, $\tau_\mathrm{f}$ and $\tau_\mathrm{s}$\newline
        Current standard deviation, $\sigma_n$\newline
        Set of evaluated points, $\mathcal{A}_n$\newline}
        
    \KwOut{Updated minimum, $\bx_\mathrm{best}$ and $f_\mathrm{best}$\newline
        Updated counters, $C_\mathrm{f}$ and $C_\mathrm{s}$\newline
        Updated standard deviation, $\sigma_n$\newline
        Updated set of evaluated points, $\mathcal{A}_{n+1}$\newline
        Updated iteration, $n$\newline}
    
    \textit{\textbf{Update counters and current best}}:
    \SetAlgoNoLine\DummyBlock{\SetAlgoLined\eIf{$f(\bx_{n+1})<f_\mathrm{best}$}{
        $C_\mathrm{s} = C_\mathrm{s}+1, C_\mathrm{f} = 0$\;
        $\bx_\mathrm{best} = \bx_{n+1}, f(\bx_\mathrm{best}) = f(\bx_{n+1})$}{
        $C_\mathrm{s} = 0, C_\mathrm{f} = C_\mathrm{f}+1$}}
    
    \textit{\textbf{Update step size}}:
    \SetAlgoNoLine\DummyBlock{\SetAlgoLined\If{$C_\mathrm{s}>=\tau_s$}{
        $\sigma_{n+1} = 2\sigma_{n}, C_\mathrm{s} = 0$}
    
    \If{$C_\mathrm{f}>=\tau_f$}{$\sigma_{n+1} = 0.5\sigma_{n}, C_\mathrm{f} = 0$}}
        
    \textit{\textbf{Update set of evaluated points}}: $\mathcal{A}_{n+1}=\mathcal{A}_n\cup \left\lbrace \bx_{n+1}\right\rbrace$\;
    
    \textit{\textbf{Update iteration number}}: $n=n+1$
    
    \caption{\textbf{Function} update\_info()}
    \label{algo:update}
\end{algorithm}

\newpage

\bibliography{references}

\end{document}